\documentclass[aps,twocolumn,showpacs,preprintnumbers,letter]{revtex4}
\usepackage{graphicx}

\newcommand{\beq}{\begin{equation}}
\newcommand{\eeq}{\end{equation}}
\newcommand{\be}{\begin{equation}}
\newcommand{\ee}{\end{equation}}
\newcommand{\ber}{\begin{eqnarray}}
\newcommand{\eer}{\end{eqnarray}}
\newcommand{\berr}{\begin{eqnarray*}}
\newcommand{\eerr}{\end{eqnarray*}}

\newcommand{\sg}{\sigma(\omega,\vec{p})}
\newcommand{\sgh}{\sigma_H(p_0,\vec{p})}

\newcommand{\addnew}[1]{#1}
\newcommand{\old}[1]{}

\def\non{\nonumber}
\def\ti{\times}

\begin{document}
\title{Quarkonium correlators and spectral functions at zero and finite temperature}
\author{A. Jakov\'ac\email{jakovac@cern.ch}}
\affiliation{
Institute of Physics, BME Budapest, Budafoki \'ut 8, H-1111 Budapest,
Hungary
 } 
\author{P. Petreczky\email{petreczk@quark.phy.bnl.gov}}
\affiliation{
Physics Department, Brookhaven National Laboratory, Upton NY 11973, USA
}
\author{K. Petrov\email{petrov@nbi.dk}}
\affiliation{
Niels Bohr Institute, Copenhagen 2100, Denmark
}
\author{A.  Velytsky\email{vel@ucla.edu}}
\affiliation{Department of Physics and Astronomy, UCLA, 
Los Angeles, CA 90095-1547, USA}

\begin{abstract}
We study quarkonium correlators and spectral functions
at zero and finite temperature using the anisotropic Fermilab lattice 
formulation with 
anisotropy $\xi=2$ and $4$. To control cutoff effects
we use several different lattice spacings. 
The spectral functions were extracted from lattice correlators with
Maximum Entropy Method based on a new algorithm. 
We find evidence
for the survival of 1S quarkonium states in the deconfined medium
till relatively high temperatures as well as for dissolution of
1P quarkonium states right above
the deconfinement temperature.
\end{abstract}
\preprint{BNL-NT 06/44 }
\maketitle

\section{introduction}
\label{sec:intro}
The study of heavy quarkonium correlators at zero and finite temperature
is interesting for several reasons. First, quarkonia are hadrons whose
properties and structures are best understood; to the first approximation 
their properties can be described 
in terms of non-relativistic potential. Also due to their small
sizes they can provide a bridge between perturbative and non-perturbative 
QCD. On general grounds it is expected that quarkonia will melt at
temperatures somewhat higher than the deconfinement temperature
as a result of modification of inter-quark forces (color screening).
Thus
their properties at finite temperature can serve as an indicator of 
in-medium modification of inter-quark forces and help to 
understand the phenomenon of non-Abelian Debye screening in Quark 
Gluon Plasma. Furthermore, it was suggested by Matsui and Satz \cite{MS86}
that color screening at high temperature will lead to quarkonium
suppression which can serve as a signal of Quark Gluon Plasma
formation in heavy ion collisions.
Quarkonium properties at finite temperature 
have been extensively studied in potential models
\cite{MS86,karsch88,ropke88,hashimoto88,digal01a,digal01b,shuryak04,wong04,wong06,mocsy05,mocsy06,mocsy06proc,alberico,rapp}.
However, the validity of potential models at finite temperature is
doubtful. It is more appropriate to study modifications of quarkonium
properties at finite temperature in terms of spectral functions.
First principle calculations of the quarkonium spectral functions at 
finite temperature have become available only 
recently \cite{umeda02,asakawa04,datta04}.
In the calculations of charmonium spectral functions which have
been performed so far, the systematic errors were not well controlled.
Although some features of the spectral functions extracted so far have been assigned to be due to lattice
discretization effects, it has not been examined in detail
which features of the charmonium spectral functions
are physical and which are merely artifacts of the finite lattice spacing
and analysis methods.

The aim of the present paper is twofold. First, we would like to study the
properties of the quarkonium spectral functions in a large range of lattice
spacings to control systematic effects both due to finite lattice spacing
and limited number of data points. 
Second, we want to study quarkonium correlators at non-zero temperature to see
what happens to different charmonium states above deconfinement. 
Having at hand results from different lattice
spacings allows to check the cutoff dependence of these results. 
In addition to charmonium correlators we also calculate bottomonium correlators
and spectral functions. Some preliminary results of this study have been
presented in \cite{kostya_lat05,peter_lat05,jhw05,vel_hard06}.

The rest of the paper is organized as follows. In section II we discuss general
properties of quarkonium correlators at finite temperature. In section III we describe
the analysis tools for the reconstruction of the quarkonium spectral functions. In
section IV we discuss our lattice setup used in the calculation of the charmonium 
correlators. In section V we discuss charmonium spectral functions at zero temperature,
emphasizing the systematic uncertainties of the analysis. Section VI deals with the 
temperature dependence of charmonium correlators. In section VII we discuss the 
problem of reconstruction of  charmonium spectral functions at finite temperature.
Section VIII contains some results on charmonium correlators at non-zero spatial
momenta. In section IX and X we discuss our results for bottomonium correlators and
spectral functions. Finally section XI contains our conclusions. The reader not 
interested in technical details may skip sections II, III and IV.

\section{Meson correlators and spectral functions}
\label{sec.corr}

In this section we discuss the relation between the
Euclidean meson correlators and spectral functions at finite
temperature. It is straightforward to take the zero temperature limit.

Most dynamic properties of the finite temperature system are incorporated 
in the spectral function. The spectral function $\sgh$ for a given 
mesonic channel $H$ in a system at temperature $T$ can be defined 
through the Fourier transform of the real time two point functions
$D^{>}$ and $D^{<}$ or equivalently as the imaginary part of 
the Fourier transformed retarded 
correlation function \cite{lebellac},
\ber
\sgh &=& \frac{1}{2 \pi} (D^{>}_H(p_0, \vec{p})-D^{<}_H(p_0, \vec{p}))
\nonumber\\
&&
=\frac{1}{\pi} Im D^R_H(p_0, \vec{p}) \nonumber \\
 D^{>(<)}_H(p_0, \vec{p}) &=& \int{d^4 p \over (2
\pi)^4} e^{i p \cdot x} D^{>(<)}_H(x_0,\vec{x}) \label{eq.defspect} \\
D^{>}_H(x_0,\vec{x}) &=& \langle
J_H(x_0, \vec{x}), J_H(0, \vec{0}) \rangle \nonumber\\
D^{<}_H(x_0,\vec{x}) &=& 
\langle J_H(0, \vec{0}), J_H(x_0,\vec{x}) \rangle , x_0>0 \
\eer

In the present paper we study local meson operators of the form
\beq
J_H(t,x)=\bar q(t,x) \Gamma_H q(t,x)
\label{cont_current}
\eeq
with
\beq
\Gamma_H=1,\gamma_5, \gamma_{\mu}, \gamma_5 \gamma_{\mu}, \gamma_{\mu} \gamma_{\nu}
\eeq
for scalar, pseudo-scalar, vector, axial-vector and tensor channels. 
The relation of these quantum number channels to different meson states is given
in Tab. \ref{tab.channels}.

\begin{table*}
\begin{tabular}
[c]{||c|c|c||c|}\hline
$\Gamma$ & $^{2S+1}L_{J}$ & $J^{PC}$ & $u\overline{u}$\\\hline
$\gamma_{5}$ & $^{1}S_{0}$ & $0^{-+}$ & $\pi$\\
$\gamma_{s}$ & $^{3}S_{1}$ & $1^{--}$ & $\rho$\\
$\gamma_{s}\gamma_{s^{\prime}}$ & $^{1}P_{1}$ & $1^{+-}$ & $b_{1}$\\
$1$ & $^{3}P_{0}$ & $0^{++}$ & $a_{0}$\\
$\gamma_{5}\gamma_{s}$ & $^{3}P_{1}$ & $1^{++}$ & $a_{1}$\\
&&$2^{++}$&\\\hline
\end{tabular}%
\begin{tabular}
[c]{|cc|}\hline
$c\overline{c}(n=1)$ & $c\overline{c}(n=2)$\\\hline
$\eta_{c}$ & $\eta_{c}^{^{\prime}}$\\
$J/\psi$ & $\psi^{\prime}$\\
$h_{c}$ & \\
$\chi_{c0}$ & \\
$\chi_{c1}$ & \\
$\chi_{c2}$ & \\\hline
\end{tabular}
\begin{tabular}[c]{|cc|}\hline
$b\overline{b}(n=1)$ & $b\overline{b}(n=2)$\\
\hline
$\eta_b$ & $\eta_b'$ \\
$\Upsilon(1S)$ & $\Upsilon(2S)$\\
$h_b$ & \\
$\chi_{b0}(1P)$& $\chi_{b0}(2P)$\\
$\chi_{b1}(1P)$& $\chi_{b1}(2P)$\\
$\chi_{b2}(1P)$&  $\chi_{b2}(2P)$\\ 
\hline                                                                   
\end{tabular}
\caption{Meson states in different channels}
\label{tab.channels}
\end{table*}

The correlators $D^{>(<)}_H(x_0,\vec{x})$ satisfy the 
well-known Kubo-Martin-Schwinger
(KMS) condition \cite{lebellac}
\beq
D^{>}_H(x_0,\vec{x})=D^{<}(x_0+i/T,\vec{x}).
\label{kms}
\eeq
Inserting a complete set of
states and using Eq. (\ref{kms}), one gets the expansion
\ber
&
\sgh = {(2 \pi)^2 \over Z} \sum_{m,n} (e^{-E_n / T} \pm e^{-E_m / T})\times \nonumber\\ 
&
\langle n | J_H(0) | m \rangle|^2 \delta^4(p_\mu - k^n_\mu + k^m_\mu) 
\label{eq.specdef}
\eer
where $Z$ is the partition function, and 
$k^{n(m)}$ refers to the four-momenta of the state $| n (m) \rangle $.

A stable mesonic state contributes a $\delta$ function-like
peak to the spectral function:
\beq
\sgh = | \langle 0 | J_H | H \rangle |^2 \epsilon(p_0)
\delta(p^2 - m_H^2),
\label{eq.stable}
\eeq
where $m_H$ is the mass of the state. For 
a quasi-particle in the medium one gets a smeared peak, with the width
being the  thermal width. 
As one increases the temperature the width increases 
and at sufficiently high
temperatures, the contribution from the meson state in the spectral function may 
be sufficiently broad so that it is not very meaningful to speak of it
as a well defined state any more. 
The spectral function as defined in
Eq. (\ref{eq.specdef}) can be directly accessible by high energy
heavy ion experiments. For example, the spectral function for the vector 
current is directly related to the differential thermal cross section 
for the production of dilepton pairs \cite{braaten90}:
\beq
{dW \over dp_0 d^3p} |_{\vec{p}=0} = {5 \alpha_{em}^2 \over 27 \pi^2} 
{1 \over p_0^2 (e^{p_0/T}-1)} \sigma(p_0, \vec{p}).
\label{eq.dilepton} \eeq
Then presence or absence of a bound state in the spectral function
will manifest itself in the peak structure of the differential 
dilepton rate.

In finite temperature lattice calculations, one calculates
Euclidean time propagators, usually
projected to a given spatial momentum:
\beq
G_H(\tau, \vec{p}) = \int d^3x e^{i \vec{p}.\vec{x}} 
\langle T_{\tau} J_H(\tau, \vec{x}) J_H(0,
\vec{0}) \rangle
\eeq
This quantity is an analytical continuation
of $D^{>}(x_0,\vec{p})$
\beq
G_H(\tau,\vec{p})=D^{>}(-i\tau,\vec{p}).
\eeq
Using this equation and the KMS condition           one can
easily show that $G_H(\tau,\vec{p})$ is related to the 
spectral
function, Eq. (\ref{eq.defspect}), by an integral equation
(see e.g. appendix B of Ref. \cite{mocsy06}):
\ber
G(\tau, \vec{p}) &=& \int_0^{\infty} d \omega
\sg K(\omega, \tau) \label{eq.spect} \non\\
K(\omega, \tau) &=& \frac{\cosh(\omega(\tau-1/2
T))}{\sinh(\omega/2 T)}.
\label{eq.kernel}
\eer
This equation is the basic equation for extracting the spectral
function from meson correlators. 
Methods to do this will be discussed in the next section.
Equation (\ref{eq.kernel})
is valid in the continuum. 
Formally the same spectral representation can be written for
the Euclidean correlator calculated on the lattice $G^{lat}(\tau,\vec{p})$.
The corresponding spectral function, however, will be distorted by the effect
of the finite lattice spacing. These distortions have been calculated in the
free theory \cite{karsch03,aarts05}.

\section{Bayesian analysis of meson correlators}
\label{sec:bayes}
The obvious difficulty in the reconstruction of the spectral function from
Eq. (\ref{eq.kernel}) is the fact that the Euclidean correlator is calculated
only at ${\cal O}(10)$ data points on the lattice, while for a reasonable discretization
of the integral in Eq. (\ref{eq.kernel}) we need ${\cal O}(100)$ degrees of freedom. The problem can be solved using Bayesian analysis
of the correlator, where one looks for a spectral function which maximizes the 
conditional probability $P[\sigma|DH]$ of having the spectral function $\sigma$ given
the data $D$ and some prior knowledge $H$  (for  reviews see \cite{asakawa01,lepagelat01}).
Different Bayesian  methods differ in the choice of the prior knowledge.
One version of this analysis which is extensively used in the literature is the 
{\em Maximum Entropy Method} (MEM) \cite{bryan,nakahara99}.
It has been used to study different correlation functions in Quantum Field Theory 
at zero and finite temperature
\cite{umeda02,asakawa04,datta04,asakawa01,nakahara99,yamazaki02,karsch02,karschqm02,dattalat02,asakawalat02,blumlat04,mysqm03,mylat01,hands,tueb,fiebig,sasaki}.
In this method the basic prior knowledge is the positivity of the spectral function and 
the prior knowledge is given by the Shannon - Janes entropy  
$$
\displaystyle
\old{S=\int d \omega \biggl [ \sigma(\omega)-m(\omega)-m(\omega)
  \ln(\frac{\sigma(\omega)}{m(\omega)}) \biggr]. }
\addnew{S=\int d \omega \biggl [ \sigma(\omega)-m(\omega)-\sigma(\omega)
  \ln(\frac{\sigma(\omega)}{m(\omega)}) \biggr]. }
$$
The real function $m(\omega)$ is called the default model and parametrizes all additional prior knowledge about the
spectral functions, e.g. such as the asymptotic behavior at high energy  \cite{asakawa01,nakahara99}.
For this case the conditional probability becomes
\beq
 P[\sigma|DH]=\exp(-\frac{1}{2} \chi^2 + \alpha S),
\addnew{\label{eq:PDH}}
\eeq
with $\chi^2$ being the standard likelihood function and $\alpha$ a real parameter.
In the existing MEM analysis of the meson spectral functions the Bryan's
algorithm was used \cite{bryan}.

Here we will introduce a new algorithm described below.
To maximize (\ref{eq:PDH}) we have to solve the following equation
\begin{equation}
  \frac{\delta}{\delta\sigma(\omega)}\left[\frac{1}{2} \chi^2 -
  \alpha S \right]=0.
\end{equation}
If we perform $N$ measurements at points $\tau_1,\dots, \tau_N$ with
results $\bar G_i$ ($i=1\dots N$), and the data model $G_i[\sigma]$ is
described by (\ref{eq.kernel}), then we obtain
\begin{equation}
  \label{eq:MEM_solve_1}
  \sum\limits_{i,j=1}^N \!K(\omega,\tau_i)\, C^{-1}_{ij}
  \left(G_j[\sigma]-\bar G_j\right) + \alpha\ln\frac{\sigma(\omega)}{m(\omega)}=0,
\end{equation}
where $C_{ij}$ is the correlation matrix. With the notation
\[ s_i\equiv -\frac1\alpha \sum\limits_{j=1}^N C^{-1}_{ij}\left(G_j[\sigma]-\bar
  G_j\right)\] 
we can write this equation as
\begin{equation}
  \sigma(\omega)= m(\omega)\exp\left[\sum\limits_{i=1}^N s_i K(\omega,\tau_i)\right].
\end{equation}
Since $s_i$ itself depends on $\sigma(\omega)$, this is just a
re-parametrization of the original problem. However, by substituting back this
form into (\ref{eq:MEM_solve_1}) we obtain
\begin{equation}
  \sum\limits_{i=1}^N \!K(\omega,\tau_i)\left[\sum\limits_{j=1}^N
  C^{-1}_{ij} \left(G_j[\sigma]-\bar G_j\right) + \alpha s_i\right]=0.
\end{equation}
Since the functions $K(\omega,\tau_i)$ are linearly independent, this
equation can hold only if the expressions in the square brackets are
zero. Multiplying by the correlation matrix it reads
\begin{equation}
  \label{eq:MEM_solve_2}
  \alpha\sum\limits_{j=1}^N C_{ij}s_j + G_j[\sigma]-\bar G_j =0.
\end{equation}
Since the relation $\sigma[s]$ is known one can solve these equations.

We can go even a step further by recognizing that
\begin{eqnarray}
  G_j[\sigma] =&& \int_0^\infty\!\! d\omega K(\omega,\tau_j) m(\omega)
  e^{\sum_n s_n K(\omega,\tau_n)}\nonumber\\ = &&
  \frac{\partial}{\partial s_j} \int_0^\infty\!\! d\omega\, m(\omega)\,
  e^{\sum_n s_n K(\omega,\tau_n)}\nonumber\\ =
  &&\frac{\partial}{\partial s_j} \int_0^\infty\!\! d\omega\,
  \sigma(\omega). 
\end{eqnarray}
This allows us to define a function
\begin{equation}
  U= \frac\alpha2\sum\limits_{i,j=1}^N s_i C_{ij} s_j +
  \int_0^\infty\!\! d\omega\, \sigma(\omega) - \sum_{i=1}^N \bar G_i s_i.
\end{equation}
Using this function (\ref{eq:MEM_solve_2}) can be written as
\begin{equation}
  \frac{\partial U}{\partial s_i}=0.
\end{equation}
It can also be shown that $\partial^2 U/\partial s_i\partial s_j$ is
positive definite. Therefore we reformulated the MEM problem to the
task of \emph{minimizing $U$}. This is a minimization problem in
number-of-data dimensions, which is not a more difficult task than
applying a $\chi^2$ method. So we should expect that our MEM-code runs
as fast as a $\chi^2$ code, as it is indeed the case. We used the
Levenberg-Marquant method to perform the minimization which in most cases 
is stable
enough to find the minimum.

It is worth to make connection between our method and the Bryan
algorithm \cite{bryan}. In both cases the true problem is
number-of-data dimensional -- in more dimensions the problem would be 
under-determined. To find the relevant subspace, the Bryan algorithm
uses singular value decomposition, while we find the same relevant
subspace by exact mathematical transformations. Although the method of
identifying the subspace is different, the result is the same, and in
both cases one proceeds with solving the original problem in this
restricted subspace. The advantage of the new algorithm is that 
it is more stable numerically when we  reconstruct quarkonium spectral functions
at zero temperature. We will discuss this issue in more detail in section \ref{sec.t0spf}.

To demonstrate the reconstruction power of     MEM        and also to
reveal the role of noise in the data we give some illustrative
examples. We use the simple uncorrelated noise model 
\begin{equation}
  \label{eq:noise}
  \delta G(\tau) = b\tau G(\tau),
\end{equation}
and vary $b$. 
For the given spectral function we generate mock data for 
$N_\tau=32$ time slices.
The terms proportional to $\alpha$ introduce the dependence on the
prior knowledge, 
so it is advantageous to choose for it smallest value possible;
the typical value in the analysis was $10^{-8}$--$10^{-12}$.

\begin{figure}[htbp]
  \centering
  \includegraphics[width=8.3cm]{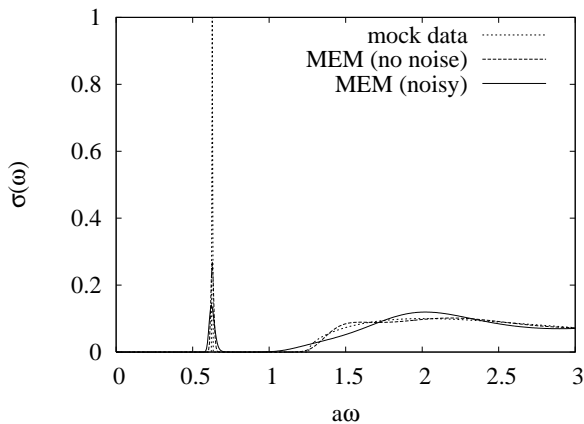}
  \caption{The continuum-like spectral function and its
  MEM-reconstruction in presence and in absence of noise,
  respectively.} 
  \label{fig:spectral_cont}
\end{figure}

The first is a typical continuum spectral function consisting of a
Dirac delta and a continuum part. In Fig.~\ref{fig:spectral_cont}
the original spectral function can be seen together with the
MEM-reconstruction in the presence and absence of noise, respectively
(the magnitude of the noise was $b=10^{-4}$). The matching is not
perfect, but all the main characteristics are present. 
The ability of MEM to reconstruct finite width spectral functions has
been demonstrated in a similar analysis with the Breit-Wigner ansatz used
to model the ground state instead of the Dirac delta \cite{vel_hard06}.
The simple noise model in this analysis assumes a diagonal correlation
matrix. In real situations the correlation matrix is not diagonal. We will
discuss this case later in the course of the analysis of the lattice correlators.

\section{Lattice formulations and parameters for charmonium physics}
\label{sec:charm_lat}

To study the charmonium system at zero and finite temperature
we use the anisotropic Fermilab formulation described in Ref. \cite{chen01}. In 
our study we use the quenched approximation and use the standard Wilson action
in the gauge sector for which the relation between the bare $\xi_0$ and the renormalized anisotropy 
$\xi=a_s/a_t$ is known in a wide range of the gauge coupling $\beta=6/g^2$ \cite{klassen}. 
For the heavy quarks we use the anisotropic clover action \cite{chen01}
\ber
&
\displaystyle
S_q^{\xi}
  = \sum_{x} {\bar{\psi}}(x) \left[
      {m_0} +
      \nu_t  \!\not\! {D}^{\rm Wilson}_t +
      \frac{\nu_s}{\xi_0}  \sum_s \!\not\! {D}^{\rm Wilson}_s  \right. \non\\
&
\displaystyle
     \left.
      - \frac{1}{2} \left(
      C_{\rm sw}^t  \sum_{s} \sigma_{ts} {F}_{ts} +  
      \frac{C_{\rm sw}^s}{\xi_0}
      \sum_{s<s\prime} \sigma_{ss\prime} {F}_{ss\prime} \right) 
      \right]  {\psi}(x) .
\label{s_fermilab}
\eer
Here the Dirac operator is defined as 
\begin{equation}
D_{\mu}^{\rm Wilson} =  \nabla_\mu - {1 \over 2} \gamma_\mu \Delta_\mu
\end{equation}
with
\begin{eqnarray}
&
 \nabla_\mu \psi(x)  = \nonumber\\
& 
{1\over 2}\, \biggl[ U_\mu(x) \psi(x+\mu) - U_{\mu}^{\dagger}(x-\mu)
           \psi(x-\mu)\biggr] \non\\
&
\\
&
 \Delta_\mu \psi(x)  = \nonumber\\
&
  \, \biggl[ U_\mu(x) \psi(x+\mu) + U_{\mu}^{\dagger}(x-\mu) \psi(x-\mu)
                                       -2 \psi(x) \biggr]   \, .\non\\
\end{eqnarray}
Furthermore, $\sigma_{\mu,\nu}=\left \{ \gamma_{\mu},  \gamma_{\nu} \right \}$ and the field strength tensor is defined as 
\ber
&
F_{\mu\nu}(x)    = 
                  - \frac{i}{2}[Q_{\mu\nu} - Q_{\mu\nu}^\dagger]  \\
&
4 Q_{\mu\nu}(x)  = 
                   U_{\mu}(x) U_{\nu}(x+\hat\mu) U_{\mu}^\dagger(x+\hat\nu)
                   U_{\nu}^\dagger(x)  \nonumber +          \non \\
               &  U_{\nu}(x) U_{\mu}^\dagger(x-\hat\mu+\hat\nu)
                   U_{\nu}^\dagger(x-\hat\mu) U_{\mu}(x-\hat\mu) \nonumber +
                                                          \non \\
               &  U_{\mu}^\dagger(x-\hat\mu) U_{\nu}^\dagger(x-\hat\mu-\hat\nu)
                   U_{\mu}(x-\hat\mu-\hat\nu) U_{\nu}(x-\hat\nu) \nonumber +
                                                          \non \\
               &  U_{\nu}^\dagger(x-\hat\nu) U_{\mu}(x-\hat\nu)
                   U_{\nu}(x+\hat\mu-\hat\nu) U_{\mu}^\dagger(x)
                                                          \non . \\
&
\eer

As in Ref. \cite{chen01}
we fix $\nu_s=1$ and use tadpole improved values for the clover coefficients. The remaining 
parameters of the action, the bare heavy quark mass $m_0$ and the bare velocity of light $\nu_t$ are fixed
non-perturbatively.    The bare quark mass is fixed by requiring that either the pseudo-scalar
($\eta_c$) mass or the spin averaged 1S mass are equal to the experimental value. 
To fix $\nu_t$ we study the dispersion relation $E(p)$ of $\eta_c$ and require that the velocity
of light defined by \cite{chen01}
\ber
&
\displaystyle
c=\frac{\sqrt{16 E^2(p_1)-15 E^2(p_0)-E^2(p_2)}}{\sqrt{12} \Delta p}\nonumber\\
&
\displaystyle
p_n=2 \pi n/N_s, \Delta p=2 \pi /N_s
\eer
is equal to one.
Here $N_s$ is the spatial size of the lattice.
For crosscheck we also calculate the velocity of light given
by the  simpler definition
\beq
c=\frac{\sqrt{E^2(p_1)-E^2(p_0)}}{\Delta p}.
\eeq
Both definitions give consistent results.
In our study we use $\xi=2,~4$ and $\beta=5.7,~5.9,~6.1,~6.5$ corresponding
to temporal lattice spacings $a_t^{-1}=1.905-14.12$ GeV.  
The bare anisotropy $\xi_0$ for a given renormalized anisotropy was calculated in Ref. 
\cite{klassen} for $\beta=5.5-6.5$. 
To set the scale for the lattice spacing we use the tradition phenomenological value $r_0=0.5$ fm for the Sommer scale \cite{sommer}.
The Sommer scale $r_0$ has also been  calculated for anisotropic
Wilson action for $\beta=5.5-6.1$ \cite{klassen_unpub}.  For larger $\beta$ values we use extrapolation based on 
$RG$ inspired $\beta$ function \cite{alton94}
\beq
\frac{a_s}{r_0}(\beta)=
\frac{1}{c_0} R(\beta) \biggr ((1+c_2 \biggr[ \frac{R(\beta)}{R(6)} \biggl ]^2  +c_4  \biggr[ \frac{R(\beta)}{R(6)} \biggl ]^4 \biggr),
\label{alton}
\eeq
with $R(\beta)=(b_0 \beta)^{-b1/(2 b_0^2)} \exp(-1/(2 b_0) \beta)$ being the standard 2-loop beta function.
The extrapolation is done for $\beta=6.5$ using $\beta=5.9,6.0$ and $6.1$.  To crosscheck
the results we also did the extrapolation including the point at $\beta=5.8$ and the difference gives the estimate
of the error on the extrapolated $r_0$. 
Alternatively one can estimate the lattice spacing from the difference between the mass of $^1P_1$ state and
the spin averaged $1S$ mass: $\Delta M(^1P_1-\overline{1S})$. To a very good approximation this
mass difference  is not affected by fine and hyperfine splitting and thus is
not very sensitive to quenching errors.
It was found that close to the continuum limit  the lattice spacing determined from  $\Delta M(^1P_1-\overline{1S})$
is different from that determined from $r_0$ by $10\%$ \cite{chen01,okamoto02} if we use the phenomenological
value $r=0.5$ fm. Using the value of $r_0=0.469(7)$ determined in full QCD \cite{gray} would give a value for 
 $\Delta M(^1P_1-\overline{1S})$ splitting which is closer to the experimental one, however, the $\Delta M(\overline{2S}-\overline{1S})$ splitting
would be even further away from the experimental value \cite{okamoto02}.
This problem is  due to the quenched approximation.
The parameters of simulations are summarized  in Tab. \ref{tab.parcc}
For $\xi=2$ they are identical to those of Ref. \cite{chen01}.
For $\xi=4$ the values of the clover coefficients have been taken from Ref. \cite{liao02}, 
and other two parameters, $m_0$ and $\nu_t$ were fixed non-perturbatively as described above.
\begin{table*}
\begin{tabular}{cccccc}
\hline
\hline
$\beta$              &       5.7          &        5.9          &         6.1           &       6.1            &      6.5  \\[3mm]
\hline
$N_s^2 \ti N_t$ &  $8^3\ti 64$  &  $16^3\ti 64$  & $16^3\ti 64$    & $16^3\ti 96$   &  $24^2 \ti 32 \ti 160$ \\[1mm]
$(\xi,~\xi_0)$     &  (2,1.6547)    &  (2,1.6907)     &   (2,1.7183)      &    (4,3.2108)   & (4,3.31655)     \\[1mm]
$r_0/a_s$         &   2.414(8)      &   3.690(11)    &  5.207(29)        &   5.189(21)     &  8.96(4)        \\[1mm]
$a_t^{-1}$[Gev] &   1.905         &   2.913           &  4.110               &   8.181           &  14.12          \\[3mm]
\hline
$C_{sw}^s$      &      2.138       &     1.889         &      1.7614        &     1.9463       &   1.7054            \\[1mm]
$C_{sw}^t$       &     1.3252      &    1.2055        &      1.1431        &     1.0984       &   0.9002            \\[3mm]
\hline
$m_0$              &     0.51           &    0.195          &      0.05            &      0.05          &  -0.025             \\[1mm]
$\nu_t$             &    1.01            &    1.09            &      1.12            &     1.25           &  1.19                \\[3mm]
\hline
$M(\eta_c)$[GeV] &    3.029(1) &   3.052(1)       &    2.994(2)       &    2.983(3)     & 3.031(9)           \\[1mm]
$c(0)$              &    1.000(2  )     &    0.984(3)      &    0.984(3)       &    1.002(4)     & 1.001(7)            \\[3mm]
\hline
$L_s$ [fm]         &     1.66           &   2.17             &   1.54              &    1.54             & 1.34                  \\[1mm]
configs              &     2000          &  1560             &    930              &    500             & 630                  \\[3mm]
\hline
\hline
\end{tabular}

\caption{Simulation parameters for charmonium at zero temperature. Also shown here is the mass of the 
$\eta_c$ state and the renormalized speed of light $c$.}
\label{tab.parcc}
\end{table*}

By studying the large distance fall-off of the correlators we determine the 
mass of the ground state in each quantum number channel.
The pseudo-scalar ground state masses from the fit are also given in Tab. \ref{tab.parcc}.

\section{charmonium spectral functions at zero temperature}
\label{sec.t0spf}

In this section we discuss the calculations
of the charmonium spectral functions at zero temperature for
different channels. The continuum meson current in Eq. \ref{cont_current}, 
$J_H$ is related to the lattice current as
\begin{equation}
J_H=Z_H a_s^3 \bar \psi \Gamma_H \psi,
\end{equation}
where $\psi$ is the lattice quark field in Eq. (\ref{s_fermilab}).
The renormalization constant $Z_H$ can be calculated in perturbation theory
or non-perturbatively. Unfortunately for anisotropic clover action they
have not been calculated. 
Since in this paper we are mainly interested in controlling systematic 
errors in the spectral functions and the temperature dependence of 
the charmonium correlators the exact values of the renormalization constants
are not important. Nonetheless, when we compare the spectral functions
calculated at different lattice spacing in the plots it is convenient to take into account
the effect of $Z_H$. We do this in an ad-hoc way by choosing $Z_H$ such that
the continuum part of the spectral function is roughly equal to its 
free value at large energies. The values of $Z_H$ are given in the Appendix.
In what follows we will discuss charmonium spectral functions at zero
spatial momentum $\vec{p}=0$.

\subsection{Pseudo-scalar and vector spectral functions at zero temperature}

In this subsection we discuss our results on
charmonium spectral functions obtained using MEM.
To reconstruct the spectral functions we used the algorithm described in
section \ref{sec:bayes}. We also compare the results with the Bryan algorithm
for the pseudo-scalar spectral function for $\beta=6.1$, $\xi=4$. The problem with the
Bryan algorithm is that it does not work well for charmonium correlators  if the time
extent is sufficiently large, which is the case at low temperatures; the iterative procedure
does not always converge. For instance at $\beta=6.1$, $\xi=4$ and $16^3 \times 96$ lattice 
we could get the spectral functions
using the Bryan algorithm only when using $t_{max}=24$ data points in the time direction.
With the new algorithm there is no restriction on $t_{max}$ which can be as large as $N_t/2$.
The comparison of the spectral functions obtained with the two algorithm is given
in the Appendix. 

When we analyze
the correlation function using MEM we need two inputs : the default
model $m(\omega)$ and the parameter $\alpha$. 
Thus the spectral function we get depends both on $\alpha$ and
the default model. 
We investigated the $\alpha$ dependence of the spectral functions.
For sufficiently large $\alpha$ the spectral function is very smooth 
and shows no peaks. As we decrease $\alpha$ the spectral function
shows more and more peaks. These features are also shown in the Appendix.
It has been shown that
one can construct the  probability $P[\alpha|Dm]$ of having some $\alpha$ for
given data and default model $m(\omega)$ \cite{bryan,jarrell96}.
Typically, for high statistics data this probability is a smooth function of $\alpha$ and 
has a maximum at some $\alpha=\alpha_{max}$ 
\cite{asakawa01,yamazaki02}. To eliminate the $\alpha$ dependence 
the spectral functions calculated at fixed $\alpha$ were averaged over 
alpha with the weight $P[\alpha|Dm]$ \cite{umeda02,asakawa04,datta04,asakawa01,nakahara99,yamazaki02}.
In the present analysis we simply
calculate the spectral function at $\alpha_{max}$. 
In the new algorithm used in this paper the sub-space, in which the search for the spectral function (i.e. the
maximization of the conditional probability $P[\sigma|DH]$ ) is performed, is chosen by selecting different data on
the correlator $G(t), t=0,1,2, ..., N_t/2$ which go into the analysis. Although the dimension of this sub-space can be
as large as $N_t/2$ in practice it is limited by statistics. In our analysis the dimension of this subspace
was tuned to the largest possible value giving a smooth $\alpha$-dependence of 
$P[\alpha|Dm]$ with a unique maximum. 
We use different type of default models, all of them being smooth
function of $\omega$, $m(\omega)=m_0 \omega^2$, $m(\omega)=h=const$
as well as the form given by the free spectral function calculated 
on the lattice \cite{karsch03}. To reduce the sensitivity to the lattice artifacts 
at short time separation arising from anisotropy,
in the reconstructed spectral function we do not include the data on the correlator 
for $t < \xi$ in our analysis. This was also done in Ref. \cite{asakawa04}.
It turns out, however, that
including these points does not alter significantly the results.

\begin{figure}
\includegraphics[width=8.5cm]{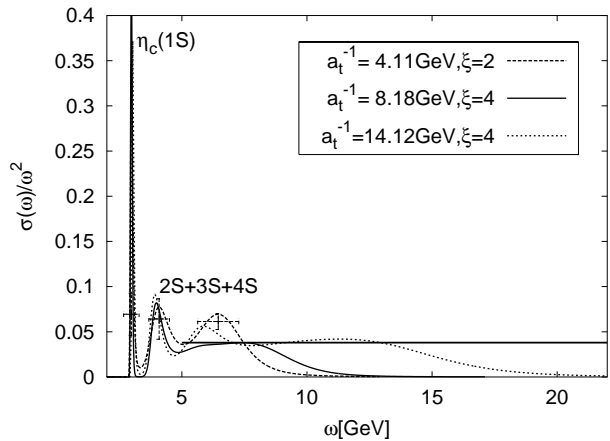}
\caption{The pseudo-scalar spectral function at zero temperature
for three finest lattice spacings. The horizontal line corresponds to the spectral function 
in the free massless limit at zero lattice spacing.}
\label{spf_ps_T0}
\end{figure}

The zero temperature spectral functions for three different lattice 
spacings are summarized in Fig. \ref{spf_ps_T0}. Here we used the
simple default model $m(\omega)=1$ (in this paper we always give the default model
in units of the spatial lattice spacing ).
To get a feeling for the statistical errors in the spectral functions we calculate
its mean value in some interval $I$:
\begin{equation} 
\bar \sigma = \frac{\int_I d \omega \sigma(\omega)}{\int_I d \omega}.
\end{equation}
Then we calculated the error on $\bar \sigma$ using standard jackknife method.
These errors are shown in Fig. \ref{spf_ps_T0}, where the length of the intervals are shown
as horizontal error bars.
As one can see from the figure, 
the $\eta_c(1S)$ can be identified very well. The second peak is likely
to correspond to excited states.
Because of the heavy quark mass the splitting between different
radial excitations is small and MEM cannot resolve different excitations
individually but rather produces a second broad peak to which all radial excitation
contribute. 
This can be seen from the fact that the amplitude of the second peak (i.e. the area under the peak)
is more than two times larger than the first one. Physical considerations tell us that it should 
be smaller than the first amplitude if it was a $2S$ state. When comparing amplitudes and 
peak positions from MEM analysis and from double exponential fits we find very good agreement for the 
first peak and a fair agreement for the second peak. The details of this comparison are given in 
the appendix. This gives us confidence that at zero temperature charmonium properties can be reproduced
well with MEM.

For energies larger than $5$GeV we probably see a continuum
in the spectral functions which is distorted by finite lattice spacing. In particular the
spectral function is zero above some energy which scales roughly as 
$a_s^{-1}$.
Note that for $\omega<5$GeV the spectral function does not depend 
on the lattice spacing.

\begin{figure}
\includegraphics[width=8.5cm]{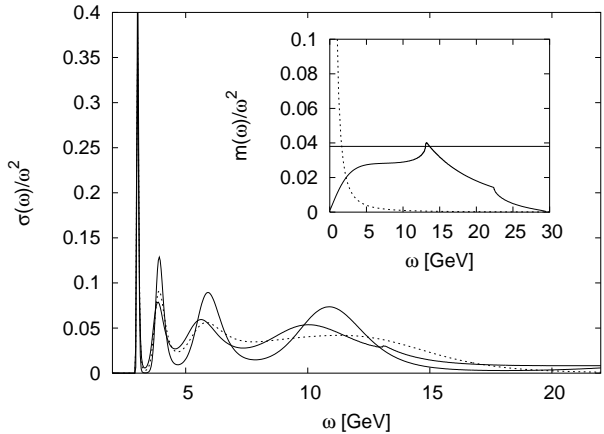} 
\caption{The default model dependence of the pseudo-scalar
spectral function at the finest lattice spacing ($\beta=6.5$).
In the insertions the default models corresponding to different
spectral functions are shown.}
\label{spf_ps_dmdep}
\end{figure}

One should control how the result depends on
the default model. In Fig. \ref{spf_ps_dmdep} we show the spectral
function for three different default models. One can see that
the default model dependence is significant only for $\omega > 5$ GeV.
This is not surprising as there are very few time slices which
are sensitive to the spectral functions at $\omega > 5$ GeV, while
the first peak is well determined by the large distance behavior
of the correlator. 

We would like to know  how well we can reconstruct the spectral
function using MEM. To answer this question we consider
a model spectral function consisting of four peaks and the
perturbative continuum. We used semi-realistic masses and amplitudes
calculated from potential model \cite{mocsy06}. 
From this spectral function we generated mock data for
the correlator $G^{mock}(i)$ at lattice spacing $a_t^{-1}=14.12$GeV and $N_t=160$. 
The correlation matrix used in the analysis of this mock data was defined as
\be
C_{ij}^{mock}=\frac{C_{ij}}{G(i) G(j)} G^{mock}(i) G^{mock}(j)
\ee
Here $C_{ij}$ and $G(i)$ are the correlation matrix and meson correlators
calculated in the pseudo-scalar channel on $24^2 \times 32 \times 160$ lattice
at $\beta=6.5$.

\begin{figure}
\includegraphics[width=8.5cm]{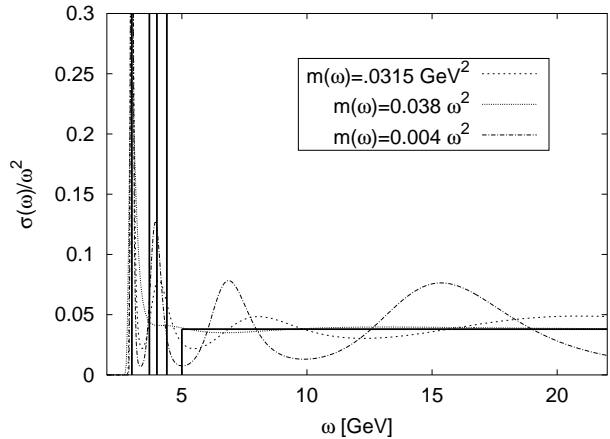}
\caption{
The comparison of spectral functions obtained from MEM with
the input spectral function (thick black lines). 
}
\label{test_mock}
\end{figure}

The result of the MEM
analysis of this mock data are shown in Fig. \ref{test_mock},
where the four delta functions and the continuum of the 
model spectral function are shown as thick black lines.
The first peak is reproduced quite well. The three radial
excitations show up as one broad peak. 
Indeed, the area under the second peak is $1.05{\rm GeV}^3$
while the input was $1.20 {\rm GeV}^3$.
The continuum is only 
reproduced correctly when the default model has exactly the same
form as the continuum at large $\omega$.

\begin{figure}
\includegraphics[width=8.5cm]{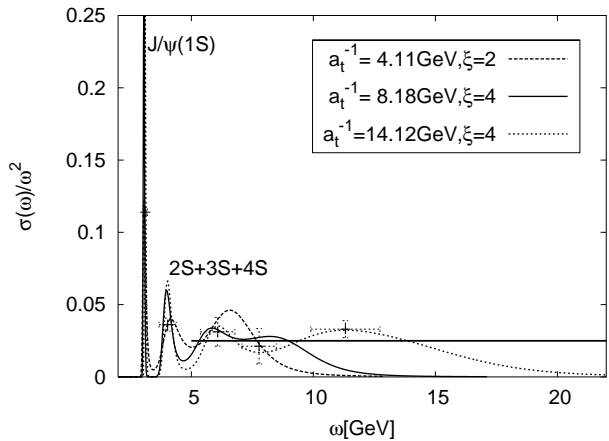}
\caption{The vector spectral function at zero temperature
for three finest lattice spacings. The horizontal line is the 
spectral function in the free massless limit.}
\label{spf_vc_T0}
\end{figure}

So far we only discussed the pseudo-scalar channel. We also
calculated the spectral function in the vector channel defined
as
\be
\sigma_V(\omega)=\frac{1}{3} \sum_i \sigma_{ii} (\omega).
\ee
The zero temperature
vector spectral functions are shown in Fig. \ref{spf_vc_T0}
for the three finest lattice spacings. 
The conclusions which can be derived from this figure are
similar to the ones discussed above for the pseudo-scalar channel.
The first peak corresponds to the $J/\psi(1S)$ state, the second peak
most likely is a combination of 2S and higher excited states, finally
there is a continuum above $5$ GeV which is, however, distorted by
lattice artifacts. The similarity between the pseudo-scalar and vector
channel is, of course, expected. The lower lying states in theses channels
differ only by small hyperfine splitting. 

\subsection{Spectral functions for P states}

\begin{figure}
\includegraphics[width=8.5cm]{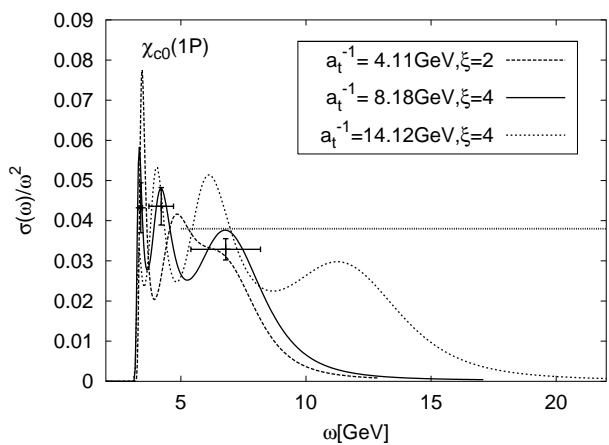}
\caption{The scalar spectral function at zero temperature
for three finest lattice spacings.}
\label{spf_sc_T0}
\end{figure}

We calculated the spectral functions in the scalar, axial-vector and
tensor channels which have the 1P charmonia  as the ground state.
The scalar spectral functions reconstructed using MEM are
shown in Fig. \ref{spf_sc_T0}.
The first peak corresponds to $\chi_{c0}$ state, but it is
not resolved as well as the ground state in the pseudo-scalar
channel. 
This is because the scalar correlator is considerably more noisy
than the pseudo-scalar or vector correlator.
For the two finest lattice spacings there is a second peak
which may correspond to a combination of excited P states.
Above $\omega>5$ GeV we see a continuum which is strongly distorted
by lattice artifacts and probably also by MEM.
In the Appendix we show the comparison of the amplitudes and the masses
of the ground state from MEM and two exponential fits. 

\begin{figure}
\includegraphics[width=8cm]{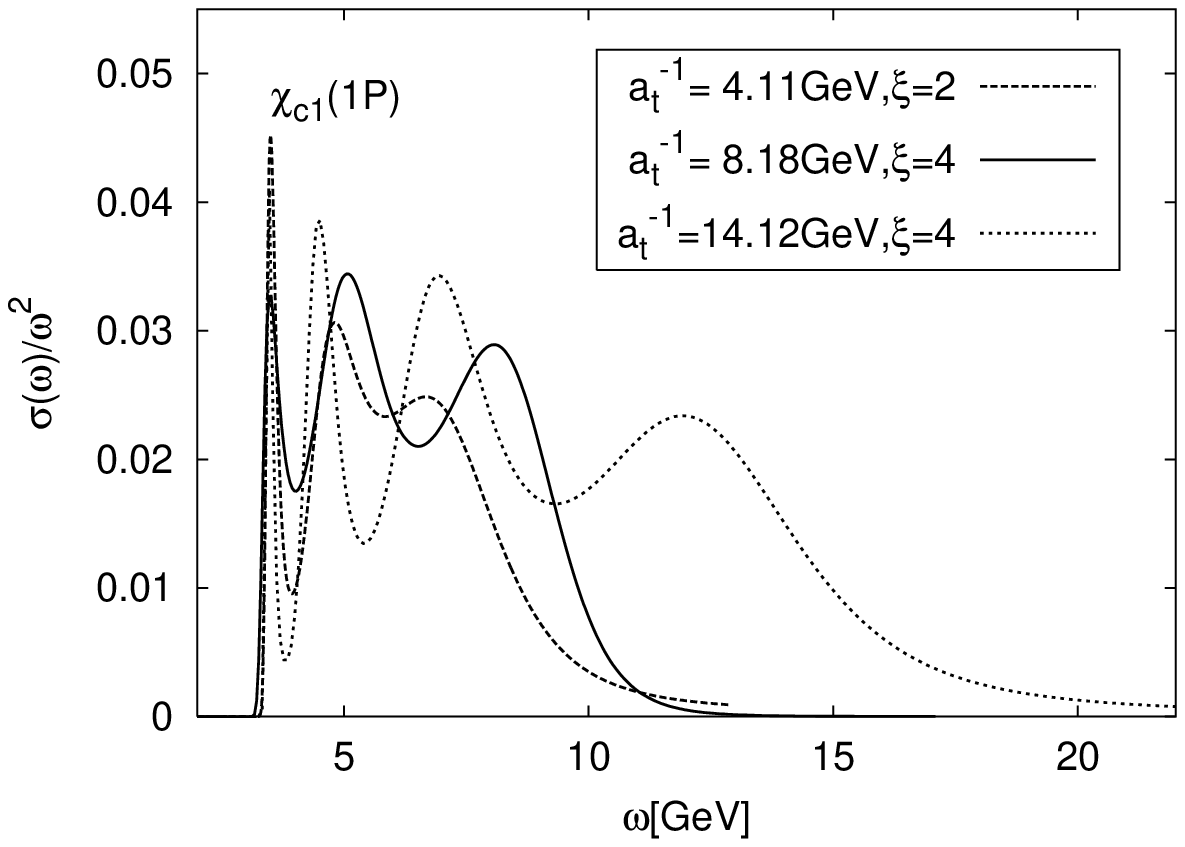}
\includegraphics[width=8cm]{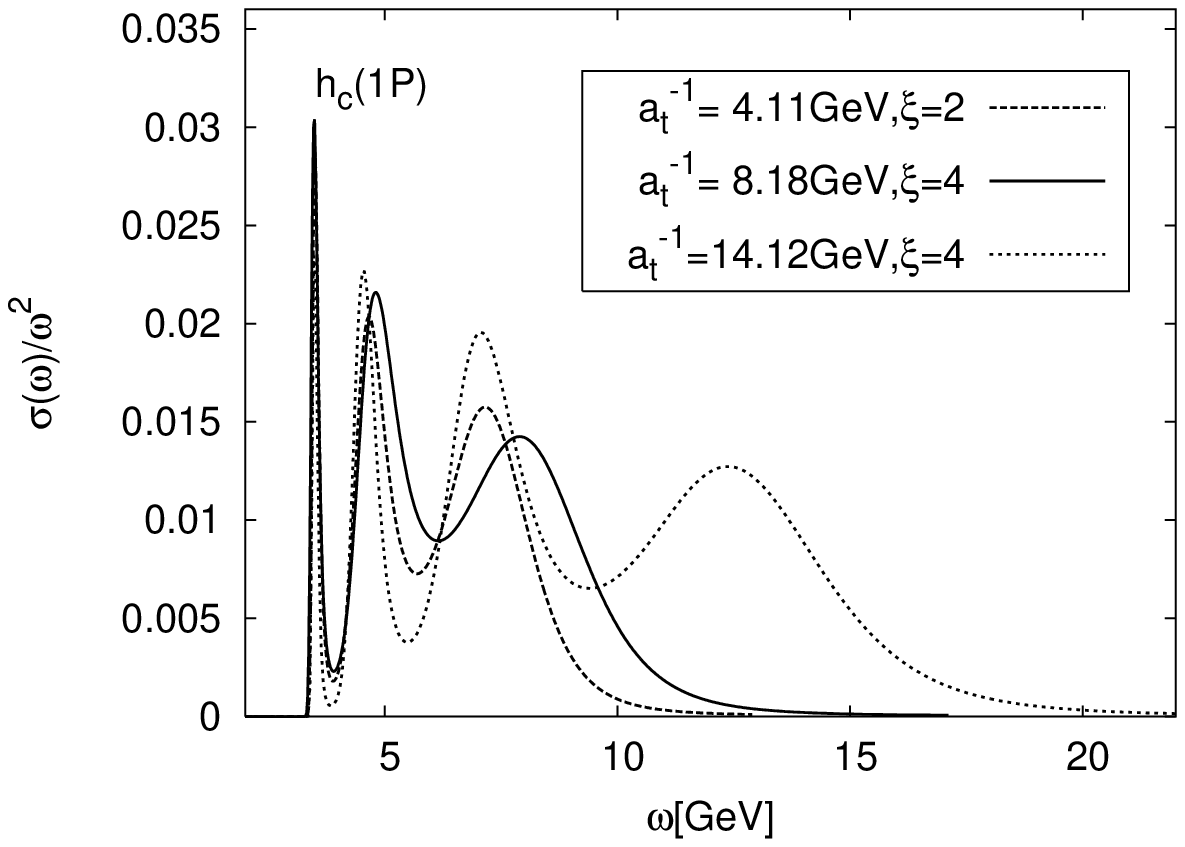}
\caption{The axial-vector (top) and tensor (bottom) spectral functions 
at zero temperature
for three lattice spacings.}
\label{spf_axt_T0}
\end{figure}

The spectral functions in the axial-vector
and tensor channels are shown in Fig. \ref{spf_axt_T0}.
They look similar to the scalar spectral functions. As in the scalar channel
the first peak is less pronounced than in the case of $S$ wave 
charmonium spectral functions, and it corresponds to $\chi_{c1}$ and 
$h_c$ state,  respectively. The continuum part of the spectral function is
again strongly distorted. 

\section{Charmonium correlators at finite temperature}

As we have seen in the previous section the reconstruction
of the spectral functions from lattice correlators is difficult
already at zero temperatures. At finite temperature it is even more 
difficult 
to control the systematic errors in the spectral functions reconstructed
from MEM. 
This is because with increasing temperature the maximal time extent
$\tau_{max}$ is decreasing as $1/T$. Also the number of data points available for the analysis
becomes smaller. 
Therefore we are looking for a method which can give
some information about the change of the spectral functions as
the temperature is increasing. The temperature dependence of the
spectral function will manifest itself in the temperature dependence
of the lattice correlator $G(\tau,T)$. Looking at Eq. (\ref{eq.kernel})
it is easy to see that the temperature dependence of $G(\tau,T)$ comes from the temperature
dependence of the spectral function $\sigma(\omega,T)$ and the 
temperature dependence of the kernel $K(\tau,\omega,T)$. To
separate out the trivial temperature dependence due to $K(\tau,\omega,T)$
following Ref. \cite{datta04} we calculate the reconstructed correlator
\begin{equation}
G_{recon}(\tau,T)=\int_0^{\infty} d \omega \sigma(\omega,T=0) K(\tau,\omega,T).
\end{equation}
If the spectral function does not change with increasing temperature 
we expect $G(\tau,T)/G_{recon}(\tau,T)=1$. In this section we are going
to study the temperature dependence of this ratio for different channels
at different lattice spacings. 
To fix the temperature scale we use the $\beta$ dependence of the $r_0$ 
given by Eq. \ref{alton} and   the value $r_0 T_c=0.7498(50)$  \cite{necco}
for the transition temperature $T_c$.
The parameters of our finite temperature
simulations are given in Table \ref{tab.ftcc}.

\begin{table*}
\begin{tabular}{|c|c|c|c|c|c|c|c|c|c|c|c|c|c|c|}
\hline
\multicolumn{3}{|c}{$\beta=5.7$, $\xi=2$} & \multicolumn{3}{|c}{$\beta=5.9$, $\xi=2$}&
\multicolumn{3}{|c}{$\beta=6.1$, $\xi=2$ } & \multicolumn{3}{|c}{$\beta=6.1$, $\xi=4$}& 
\multicolumn{3}{|c|}{$\beta=6.5$, $\xi=4$}\\
\hline
lattice     &$T/T_c$&\#config  &    lattice       &$T/T_c$&\#config    & lattice      &$T/T_c$      &\#config    &        lattice          &$T/T_c$ &\#config     &        lattice          & $T/T_c$&\#config\\
$8^3\times6$& 1.08  & 500    & $16^3\times16$& 0.62  & 1000      &$16^3\times 24$ & 0.58  & 1000      & $16^3\times 32$ & 0.87  & 220               &$24^3 \times44$  & 1.09     & 110     \\
                      &       &             & $16^3\times 8$& 1.23  & 1000       &$16^3\times 12$  & 1.16  & 250       & $16^3\times 26$ & 1.07  & 400               &$24^3 \times 40$ & 1.20     &1680 \\
                      &       &             &                         &          &                &                            &         &               & $16^3\times 24$ & 1.16  & 2010             &$24^2 \times 32 \times 32$ & 1.50     & 1000    \\
                      &       &             &                         &          &                &                            &         &               &$24^3\times  20$ & 1.39  & 1900             &$24^3 \times 24$  & 1.99     & 300    \\
                      &       &             &                         &          &                &                            &         &               &$16^3\times  16$ & 1.73  & 1000             &$24^3 \times 20$  & 2.39     & 640    \\
                      &       &             &                         &          &                &                            &         &               &$16^3\times  12$ & 2.31  & 3000             &$24^3 \times 16$  & 2.99     & 310     \\
                      &       &             &                         &          &                &                            &         &               &$24^3\times   24$ &1.16  & 2040             &                             &             &         \\
\hline
\end{tabular}
\caption{Simulation parameters for charmonium at finite temperature. We assumed $r_0 T_c=0.7498(50)$ corresponding to $T_c=295(2)$ MeV. }
\label{tab.ftcc}
\end{table*}

\subsection{The pseudo-scalar correlators}

\begin{figure}
\includegraphics[width=8.3cm]{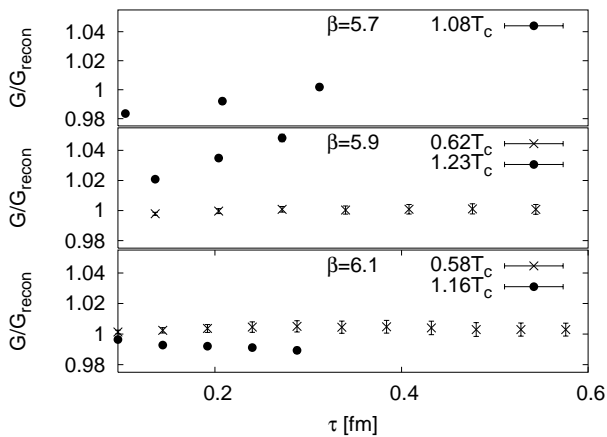} 
\caption{The ratio $G/G_{recon}$ for the pseudo-scalar channel
on coarse $\xi=2$ lattices.}
\label{psxi2}
\end{figure}

\begin{figure}
\includegraphics[width=8.3cm]{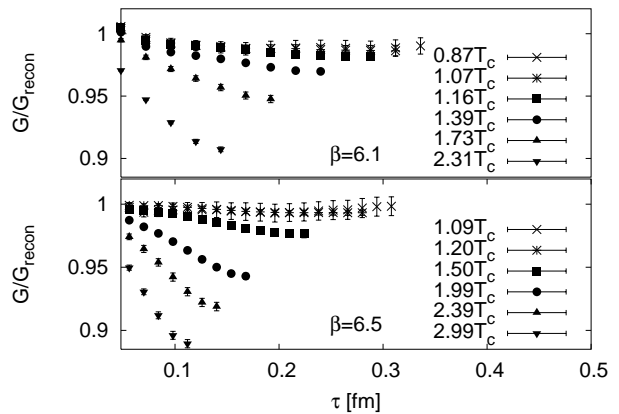} 
\caption{The ratio $G/G_{recon}$ for the pseudo-scalar channel
for the two finer $\xi=4$ lattices.}
\label{psxi4}
\end{figure}

First let us examine the temperature dependence of the pseudo-scalar
correlators. In Fig. \ref{psxi2} we show our numerical results for
$G/G_{recon}$ on coarse lattices with $\xi=2$. The figure shows very little temperature
dependence of the correlators till temperatures $1.2T_c$. 
Calculations at smaller lattice spacings enable us to consider higher
temperatures. In Fig. \ref{psxi4} we show the temperature dependence of
$G/G_{recon}$ on our $\xi=4$ lattices. 
We see very little change in the pseudo-scalar correlator till temperatures
as high as $1.5T_c$. Medium modifications of the correlator slowly
turn on as we increase the temperature above this value.
From the figures it is clear that the temperature dependence of 
the correlators is not affected significantly by the finite lattice spacing. 
The very small temperature dependence of the pseudo-scalar correlator suggests
that the corresponding ground state $\eta_c(1S)$ survives till
temperatures as high a $1.5T_c$. The temperature dependence of the
correlator found in this study is similar to that of Ref. \cite{datta04} where
isotropic lattices with very small lattice spacings, $a^{-1}=4.86,9.72$ GeV
have been used.
We find a somewhat stronger temperature dependence of $G/G_{recon}$ than
in Ref. \cite{datta04}. In particular, at $1.5T_c$ we see small, but statistically significant deviations of
$G/G_{recon}$  from unity. 
At higher temperatures the deviation of $G/G_{recon}$
from unity  become slightly larger than those found in Ref. \cite{datta04}. This is possibly due to the fact
that cutoff effects  are more important at higher temperatures. 
Thus despite similarities of the temperature dependence of the pseudo-scalar
correlator to findings of Ref. \cite{datta04} we see quantitative differences.
One should note, however, statistical errors and systematic uncertainties are larger
in the analysis presented in Ref. \cite{datta04} than in this calculation.
The ratio
$G/G_{recon}$ starts to depend more strongly on the temperature around $2T_c$. 
This may suggest some quantitative differences in the properties of the lowest state
at this temperature.

\subsection{The P-wave correlators}

\begin{figure}
\includegraphics[width=8.3cm]{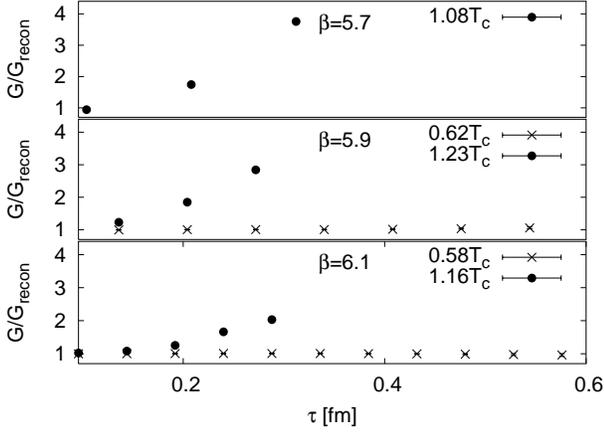}
\caption{The ratio $G/G_{recon}$ for the scalar channel
on coarse $\xi=2$ lattices.}
\label{scxi2}
\end{figure}

\begin{figure}
\includegraphics[width=8.3cm]{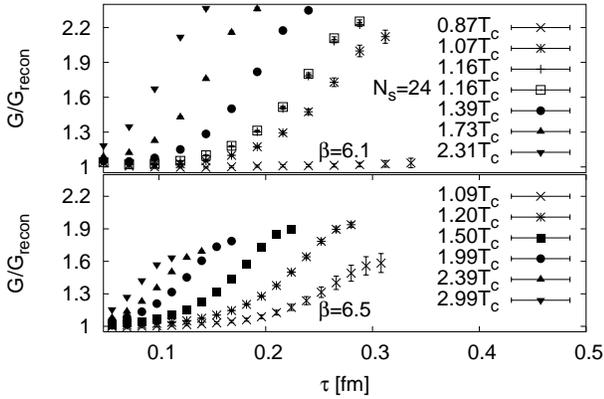}
\caption{The ratio $G/G_{recon}$ for the scalar channel
for the two finer $\xi=4$ lattices.}
\label{scxi4}
\end{figure}

In this subsection we are going to discuss the temperature dependence of
the scalar, axial-vector and tensor correlators corresponding to P-states. 
The numerical results for the scalar correlator on our $\xi=2$
coarse lattices are shown in Fig. \ref{scxi2}. As one can see
the correlator is temperature independent below $T_c$ and strongly
enhanced above $T_c$. The magnitude of the enhancement is largest
on the coarsest lattice and decreases with decreasing lattice spacing.
The numerical results on fine
lattices are shown in Fig. \ref{scxi4}. 
We see some differences in $G/G_{recon}$ calculated at $\beta=6.1$ and
$\beta=6.5$.  Thus the cutoff dependence of $G/G_{recon}$  is larger
in the scalar channel than in the pseudo-scalar one. For $\beta=6.1$ and $\xi=4$
we also did calculations on $24^3 \times 24$ lattice to check finite volume effects.
The corresponding results are shown in Fig. \ref{scxi4} indicating that the finite
volume effects are small.
On the finest lattice
the enhancement of the scalar correlator is very similar to that
found in calculations done on isotropic lattices \cite{datta04}, but small
quantitative differences can be identified.

\begin{figure}
\includegraphics[width=8.3cm]{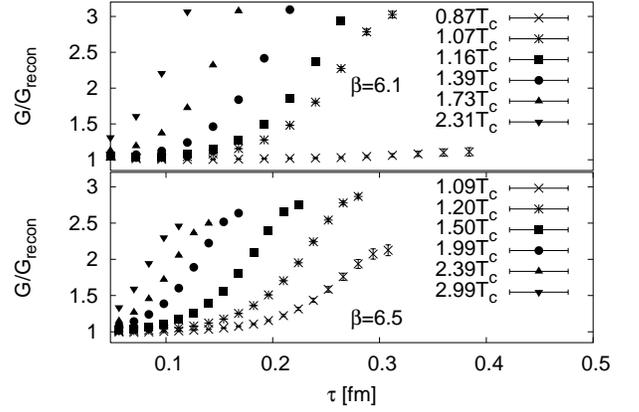}
\caption{The ratio $G/G_{recon}$ for the axial-vector channel
for $\xi=4$ lattices.}
\label{axxi4}
\end{figure}

\begin{figure}
\includegraphics[width=8.3cm]{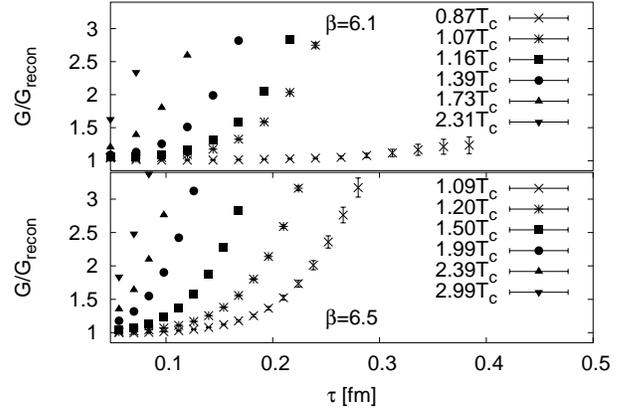}
\caption{The ratio $G/G_{recon}$ for the tensor channel
for $\xi=4$ lattices.}
\label{b1xi4}
\end{figure}

In Figs. \ref{axxi4} and \ref{b1xi4} we show the temperature dependence of the
axial-vector and tensor correlators respectively for $\xi=4$.
Qualitatively their behavior is very similar to the scalar correlator but the enhancement
over the zero temperature result is larger. The results for the axial-vector correlators 
again are very similar to those published in Ref. \cite{datta04}.
The difference in $G/G_{recon}$ calculated  at $\beta=6.1$ and $\beta=6.5$ are
smaller than in the scalar channel.

The large increase in the scalar, axial-vector and tensor correlators 
indicates strong modification of the corresponding spectral function and, possibly the dissolution of 1P charmonia
states.

\subsection{The vector correlator}

\begin{figure}
\includegraphics[width=8.3cm]{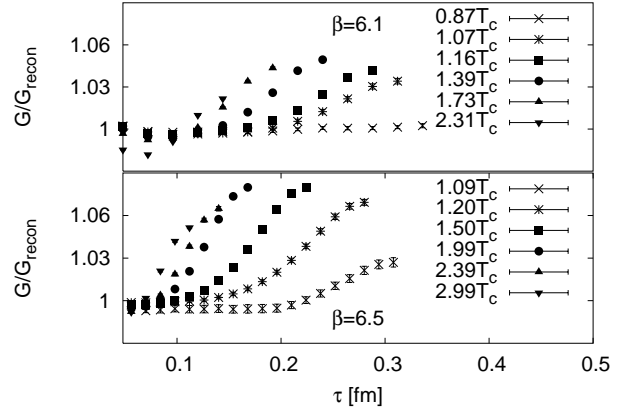}
\caption{The ratio $G/G_{recon}$ for the vector channel
for the two finer $\xi=4$ lattices.}
\label{vcxi4}
\end{figure}

\begin{figure}
\includegraphics[width=8.3cm]{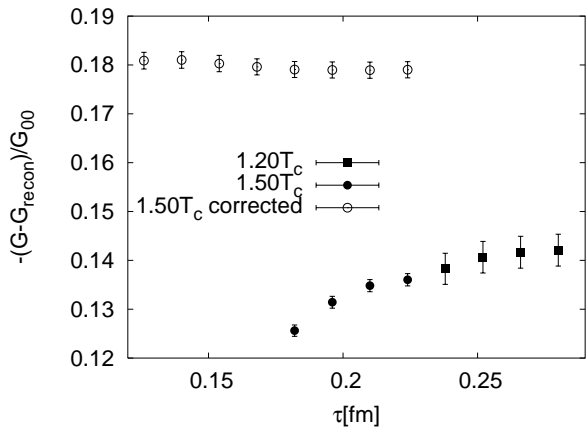}
\caption{The ratio $(G-G_{recon})/G_{00}$ for the vector channel
at $1.2T_c$ and $1.5T_c$. The open symbols show the same quantity at
 $1.5T_c$ where the possible modifications of the high energy part of the 
finite temperature spectral functions have been taken into account (see text).
} 
\label{fig:TpM}
\end{figure}

The numerical results on the vector correlator are shown
in Fig. \ref{vcxi4}
for $\xi=4$. As one can see from the figures the temperature dependence of $G/G_{rec}$ is
different from the pseudo-scalar case and this ratio is larger than unity for all lattice spacings.
Though this appears to be surprising 
at first glance it can be explained by the presence of the diffusion pole in the vector 
spectral function \cite{mocsy06,derek}. Since the vector current is conserved there is a 
low energy contribution in the vector spectral function at finite temperature
\begin{eqnarray}
&
\sigma^V_{ii}(\omega \ll T) =\chi_s(T) \langle v_{th}^2 \rangle  \omega \delta_{\eta}(\omega)
\label{sigmaii}\\
&
\sigma^V_{00}(\omega \ll T)=-\chi_s(T) \omega \delta(\omega),
\label{sigma00}
\end{eqnarray}
where $ \langle v_{th}^2 \rangle $ is the average thermal velocity of the heavy quark.
For free streaming heavy quarks at temperature $T$ we have $ \langle v_{th}^2 \rangle =T/M$,
with $M$ being the heavy quark mass.
Furthermore, 
$\chi_s(T)$ is the quark number susceptibility 
corresponding to the heavy  quark, which in the non-interacting case is given by:
\begin{equation}
\chi_s(T)=\frac{1}{T} \int \frac{d^3 p}{(2 \pi)^3} \frac{1}{\exp((\sqrt{p^2+M^2})/T)+1}.
\end{equation}
The low energy parts of $\sigma^V_{00}$ and  $\sigma^V_{ii}$
correspond to charge fluctuations and heavy quark diffusion, respectively. The smeared delta function, $\delta_{\eta}(\omega)$,
in Eq.  (\ref{sigmaii}) carries information about the heavy quark diffusion constant $D$ \cite{derek}.
Using effective Langevin theory \cite{derek} it can be shown that 
\begin{equation}
\delta_{\eta}(\omega)=\frac{1}{\pi} \frac{\eta}{\omega^2  + \eta^2},~~\eta=\frac{T}{M D}.
\end{equation}
Therefore a very precise calculation of the vector correlator at finite temperature
can provide some information about heavy quark transport in Quark Gluon Plasma.
This would require a method which is capable to isolate reliably the low energy contribution given
to the Euclidean correlator which is given by Eq. (\ref{sigmaii}) .
In the present paper we attempted to do so by subtracting from
the finite temperature spectral functions the corresponding high energy part. We have assumed that this
high energy part can be approximated reasonably well by the zero temperature spectral function.
As we have seen in the previous subsection the pseudo-scalar correlators show little temperature
dependence up to temperatures as high as $1.5T_c$. This suggests that to some extent the spectral
function up to this temperature can be approximated by the zero temperature spectral function. 
Since the lowest lying states in the pseudo-scalar and vector channel are the same (up to the small hyperfine splitting) we may expect
that the same is true for the vector channel,  and that the only difference is the transport peak at $\omega \simeq 0$.
Therefore we expect that $G^V-G^V_{recon}$ should give an estimate for the low energy contribution coming from Eq. (\ref{sigmaii}).
It is easy to see that for the temporal component of the vector correlator we have $G_{00}=-T \chi_s(T)$ ( c.f. Eq. (\ref{sigma00})).
This is a consequence of charge conservation. In Fig. \ref{fig:TpM} we show $-(G^V(\tau)-G_{recon}^V(\tau))/G_{00}$ as function of $\tau$.
Note that this quantity does not depend on the renormalization constant.
Since the width of the smeared delta function in Eq. (\ref{sigmaii}) is small the above quantity should be approximately independent of
$\tau$ and give the averaged thermal velocity    $ \langle v_{th}^2 \rangle $.
The data in Fig. \ref{fig:TpM} do not agree completely with these expectations, deviations from this behavior are seen 
at short distances. The deviations are significant for $1.5T_c$.
The problem is that with increasing temperature the high energy part of the spectral function
starts to deviate from the zero temperature result, e.g. the properties of the lowest state could be slightly modified.
The difference  $G^V(\tau)-G_{recon}^V(\tau)$ can be quite sensitive to these small changes, especially as the temperature 
increases. 

It has been noticed already in the previous subsection that at $1.5T_c$ we see small but significant temperature
modifications of the ratio $G/G_{recon}$ for the pseudo-scalar channel. Therefore at this temperature we attempted to
take into account possible modifications of the high energy part of the spectral function. We do this by lowering 
the amplitude of the 1st peak by $7\%$ in the zero temperature spectral function. 
In the pseudo-scalar channel this ensures that $G/G_{recon}$  stays around unity. Using this modified
spectral function in the vector channel we construct the quantity $-(G^V(\tau)-G_{recon}^V(\tau))/G_{00}$ again.
This is also shown in Fig. \ref{fig:TpM} as open symbols. We see that as a result of this correction
 $-(G^V(\tau)-G_{recon}^V(\tau))/G_{00}$ shows a nice plateau and also increases with the temperature as 
expected. For the averaged thermal velocity squared we estimate  $ \langle v_{th}^2 \rangle \simeq 0.13$ for $1.2T_c$ and
$ \langle v_{th}^2 \rangle \simeq 0.18$ for $1.5T_c$.  A non-zero diffusion coefficient $\eta$ would 
produce a small curvature in $-(G^V(\tau)-G_{recon}^V(\tau))/G_{00}$, clearly within present 
systematic uncertainties we cannot make any statement on the value of $\eta$.

The quantity $(G-G_{recon})/G_{00}$ gives an estimate of $T/M$ 
and is shown in Fig. \ref{fig:TpM}. As one can see from the figure it has the
expected magnitude.

\section{Charmonium spectral functions at finite temperature}

The study of the temperature dependence of the charmonium 
correlators presented in the previous section provides some useful
insight into the properties of the spectral functions in the deconfined 
phase. In particular it suggests the survival of the 1S state till
temperatures as high as $1.5T_c$ and significant modification or dissolution of the
1P charmonium states. To get more detailed information about charmonium 
spectral functions we would like to use MEM.  
We attempted to reconstruct the charmonium spectral functions at finite
temperature on our two finest lattices.
Below we are going to present
our results for charmonium spectral functions at finite temperature for
different channels.

\subsection{Pseudo-scalar and vector spectral functions at finite temperature}

\begin{figure}
\includegraphics[width=8.6cm]{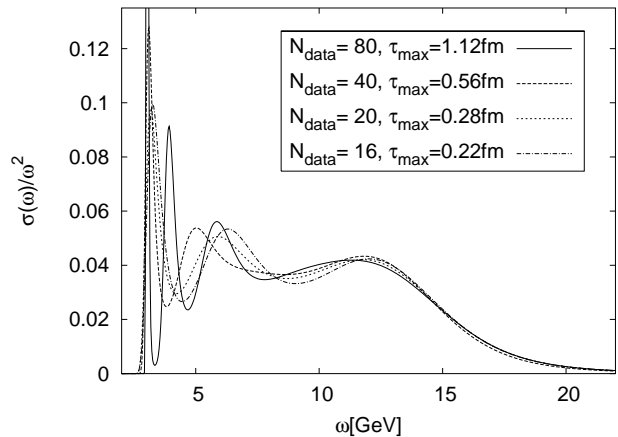}
\caption{The dependence of the reconstructed pseudo-scalar spectral function on the maximal
temporal extent for $\beta=6.5$. In the analysis the default model $m(\omega)=1$ has been
used.
}
\label{spf_ps_ntdep}
\end{figure}

In section \ref{sec.t0spf} we have seen that using MEM we can 
reconstruct well the main features of the spectral function, in particular
the ground state properties. At finite temperature the situation becomes
worse because the temporal extent is decreasing. The maximal time 
separation is $\tau_{max}=1/(2T)$. As a consequence 
it is no longer possible to isolate the ground state  well. 
Also the number of available data points becomes smaller.  While the later 
limitation can be overcome by using finer and finer lattice spacing in time direction
the former limitation is always present. Therefore we should investigate 
systematic effects due to limited extent of the temporal direction. It appears
that the pseudo-scalar channel is the most suitable case for this investigation
as at zero temperature it is well under control and there is no contribution from
heavy quark transport.
To estimate the effect of limited temporal extent we have calculated the 
spectral functions at zero temperature considering only the first $N_{data}$ 
time-slices in the analysis for $\beta=6.5$, $\xi=4$. 
The result of this calculation is shown in Fig. \ref{spf_ps_ntdep}
where we considered $N_{data}=80,~40,~20$ and $16$.  The last
two values correspond to our finite temperature lattices at this value of $\beta$.
In this case we see the first peak quite clearly when $\alpha=\alpha_{max}$.
For $\beta=6.1$ as well as for $\beta=6.5$ and $T>1.5T_c$  the 1st peak is only clearly visible when we consider
values for $\alpha$ which are smaller  than $\alpha_{max}$.
As one can see from the figure already for $N_{data}=40$ and $\tau_{max}=0.56$ fm
the second peak corresponding to radial excitation is no longer visible and the first peak
becomes significantly broader. The position of the first peak, however, is unchanged.
As the number of data points is further decreased ($N_{data}=20,16$) we see further broadening of the first peak
and a small shift of the peak position to higher energies.   
These systematic effects should be taken into account when analyzing the spectral functions
at finite temperature. Therefore when studying the spectral functions at finite temperature
we always compare with the zero temperature spectral functions reconstructed 
with the same number of data points and $\tau_{max}$ as available at that temperature.

\begin{figure}
\includegraphics[width=8.6cm]{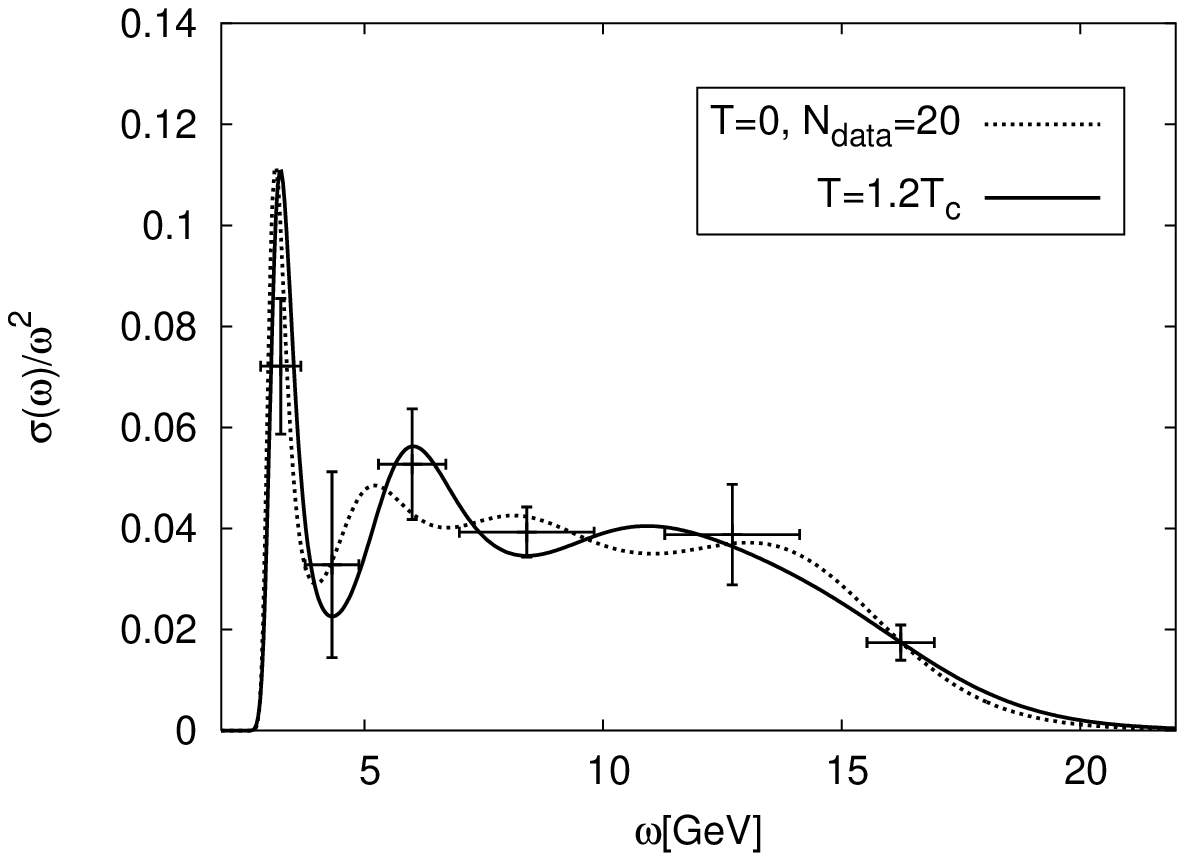}
\includegraphics[width=8.6cm]{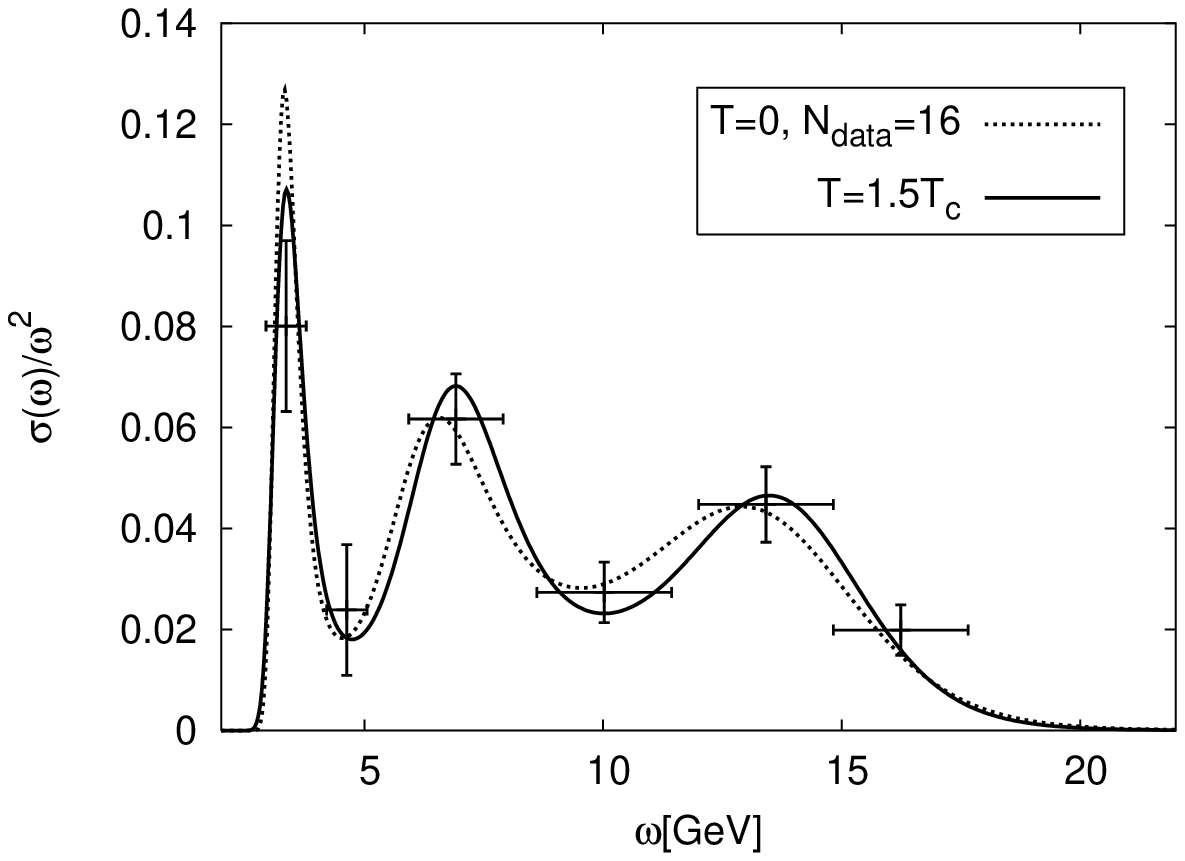}
\includegraphics[width=8.6cm]{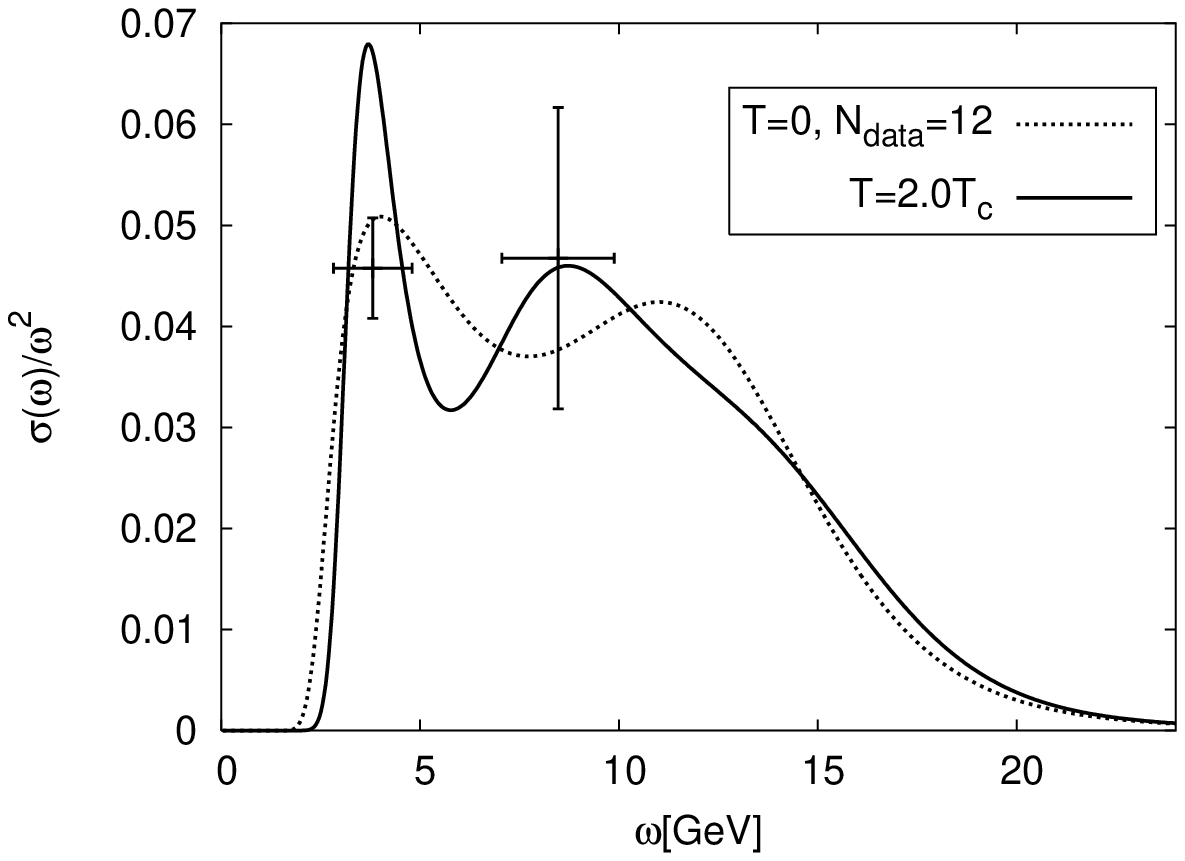}
\caption{The pseudo-scalar spectral function
for $\beta=6.5$ and $N_t=40,~32,20$ corresponding to temperatures
$1.2T_c,~1.5T_c$ and $2.4T_c$. 
In the analysis the default model $m(\omega)=0.01$ has been
used for $1.2T_c$ and $1.5T_c$, while for $2.0Tc$ we used $m(\omega)=1$.
}
\label{spf_ps_Tne0}
\end{figure}

In Fig. \ref{spf_ps_Tne0} the spectral functions at different temperatures are shown
together with the zero temperature spectral functions. We used $m(\omega)=0.01$ as a default model
for $T=1.2T_c$ and $T=1.5T_c$. For $T=2.0T_c$ we used  $m(\omega)=1$ since the use of  $m(\omega)=0.01$ 
resulted in multiple maxima for $P[\alpha|DH]$.
The figure shows that the pseudo-scalar spectral function is not modified till $1.5T_c$ within the
errors of our calculations.
This is consistent with the conclusions of Ref. \cite{asakawa04,datta04}.
One should note, however, that it is difficult to make any conclusive statement based on the shape
of the spectral functions as this was done in the above mentioned works. The dependence of
the reconstructed spectral functions on the default model $m(\omega)$ is much stronger
at finite temperature. We have reconstructed the spectral functions using different type of default models.
For all 
temperatures  $T \le 1.5T_c$ the difference between the finite temperature spectral function
and the zero temperature one is very small compared to the statistical errors for all default model considered here. 
We discuss this further in the Appendix.
In particular we used the default model constructed from the high energy part
of the lattice spectral functions calculated at zero temperature as this was done in Ref. \cite{datta04}.
The idea is that at sufficiently high energy the spectral function is dominated by the continuum and 
is temperature independent and is suitable to provide the prior knowledge, i.e. the default model. This 
behavior of $m(\omega)$ is then matched to a simple $\omega^2$ dependence \cite{datta04}. With this
default model we have calculated the spectral function also at $\beta=6.1$ on $16^3 \times 26$ and
$16^3 \times 24$ lattices corresponding to $1.07T_c$ and $1.16T_c$. Also here we have found very
little temperature dependence of the spectral functions. The survival of the 1S charmonium state was
also confirmed by the analysis of the correlation functions with different spatial boundary condition
\cite{doi}. 

At sufficiently high temperatures $1S$ charmonium states should dissolve and thus we expect
significant changes in the spectral functions. The analysis of the pseudo-scalar correlators suggests
that this may happen around $2T_c$ which was also the conclusion of Ref. \cite{asakawa04}.
The spectral function at  $T=2.0T_c$ shown in Fig. \ref{spf_ps_Tne0} shows significant
deviation from the zero temperature spectral function. However, this conclusions could be
premature, since for a different default model, namely the free lattice spectral function,  we see almost
no temperature dependence of the spectral function also at $2T_c$. Using the high energy part of the zero temperature spectral
function as the default model as discussed above we arrive at a similar conclusion. Thus further studies are needed
to establish at which temperature the $1S$ charmonia states dissolve.

\begin{figure}
\includegraphics[width=8.6cm]{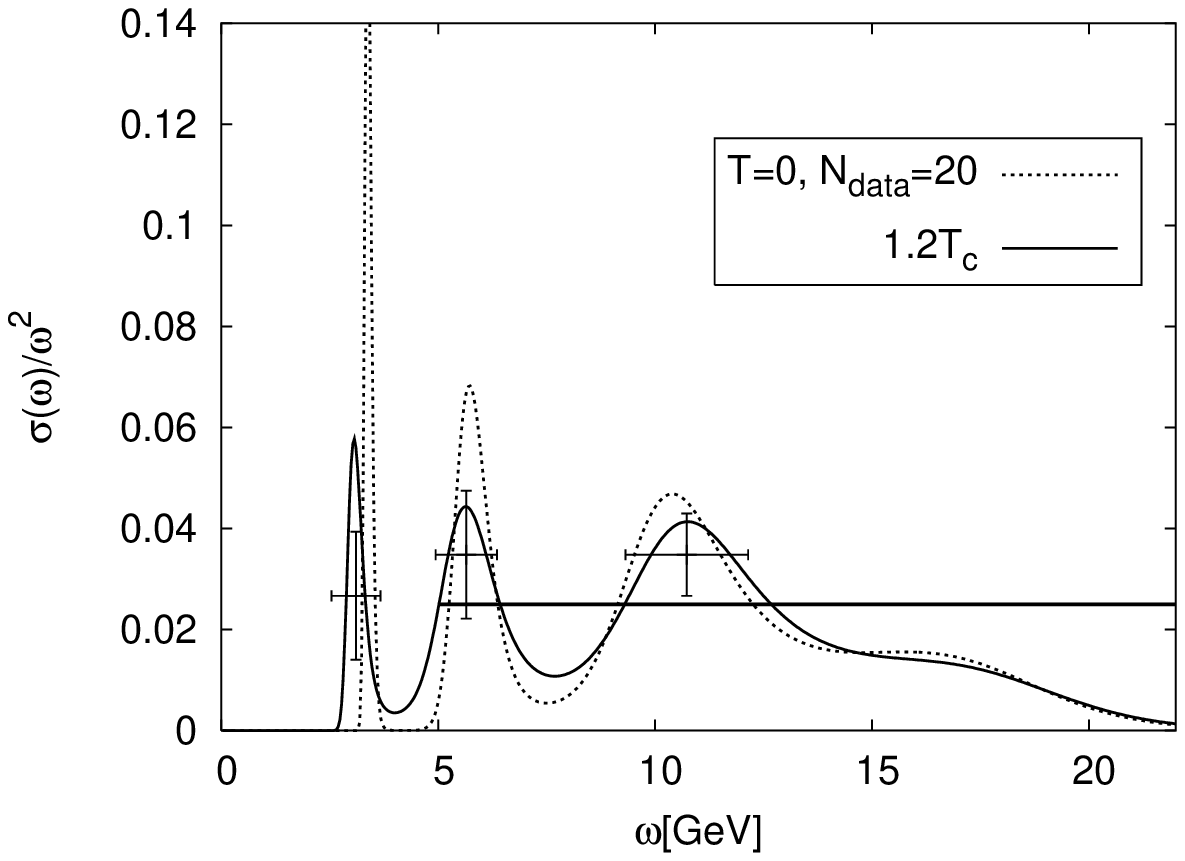}
\includegraphics[width=8.6cm]{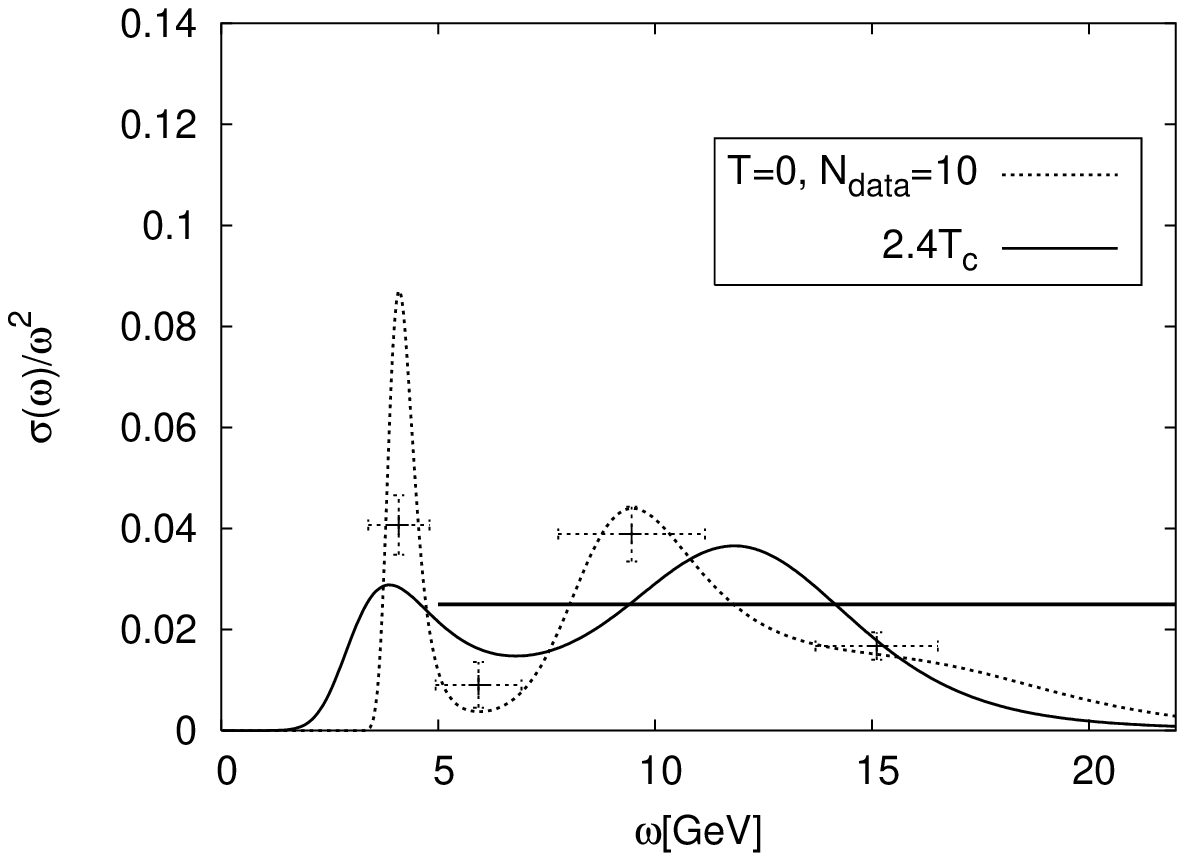}
\caption{The vector spectral function
for $\beta=6.5$ and $N_t=40,~20$ corresponding to temperatures
$1.2T_c$ and $2.4T_c$. 
In the analysis the default model $m(\omega)=0.01$ has been
used.
}
\label{spf_vc_Tne0}
\end{figure}

We also calculated the spectral function in the vector channel. The results are shown in Fig. \ref{spf_vc_Tne0}
for the default model $m(\omega)=0.01$ .
As this was already discussed in the previous section the basic difference between the pseudo-scalar
and vector spectral functions at finite temperature is the presence of the transport peak at $\omega\simeq 0$.
The difference of the temperature dependence of the vector and pseudo-scalar correlators is consistent
with this assumption. The vector spectral function reconstructed with MEM shows no evidence of the transport
peak at $\omega \simeq 0$. On the other hand the spectral function at $1.2T_c$ differs from the zero temperature
spectral function, in particular the first peak is shifted to smaller $\omega$ values. We believe this is a problem of the MEM
analysis
which cannot resolve the peak at $\omega \simeq 0$ but instead mimics its effect by shifting the $J/\psi$ peak to smaller
$\omega$.  Also at $2.4T_c$ the spectral function extends to smaller $\omega$ values than in the pseudo-scalar
correlator which again indicates some structure at $\omega \simeq 0$. We analyzed the vector spectral functions
using other choices for the default model and always found that the spectral functions at finite temperature differs from
the zero temperature spectral functions and extend to significantly smaller $\omega$ values.

\subsection{The P-wave spectral functions}

\begin{figure}
\includegraphics[width=8.6cm]{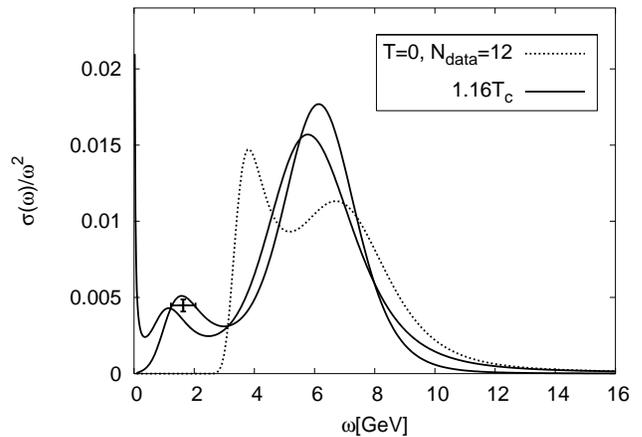}
\caption{The scalar spectral function for $\beta=6.1$ at $T=1.16T_c$
and at zero temperature reconstructed using $N_{data}=12$. 
At finite temperature two default models $m(\omega)=0.01$ and $m(\omega)=0.038 \omega^2$ have been
used.
}
\label{spf_sc_Tne0_1}
\end{figure}

We have studied the spectral functions in the scalar, axial-vector and tensor
channels. Already at zero temperature the calculations of the P-wave spectral functions
turned out to be much more difficult than for the S-wave spectral functions. This is even more the case
at finite temperature where the number of data points is limited. For most of the choices of the default model
we have found that $P[\alpha|DH]$ has multiple maxima for all of our data sets. This is a sign of insufficient
statistics. The only exception is the data set corresponding
to $16^3 \times 24$  lattice with $\beta=6.1$, i.e. $T=1.16T_c$.
The spectral function in the scalar channel for $\beta=6.1$ at $T=1.16T_c$  is shown in Fig. \ref{spf_sc_Tne0_1}
and compared with spectral functions at zero temperature reconstructed using $N_{data}=12$ data points.
At finite temperature two default models, $m(\omega)=0.038$ and $m(\omega)=0.01$ have been used. 
We see that the spectral function at finite temperature is significantly modified compared 
to the zero temperature spectral function; its shape at small $\omega$, however, strongly 
depends on the default model. We also analyzed the spectral function for these data with
the default model constructed from the high energy part of the zero temperature lattice spectral functions
as described in the previous subsection. The spectral functions obtained from this analysis are similar
to those shown in Fig. \ref{spf_sc_Tne0_1} (see Appendix).

From the analysis of the spectral functions we conclude that the 1P states are dissociated
in the deconfined phase at temperatures $T \simeq 1.1-1.2T_c$. 
This is consistent with the analysis of the correlators presented in the previous section
as well as with the conclusions of Ref. \cite{datta04}.
On the other hand with present
level of statistical accuracy we cannot make precise statements about the form of the P-wave
spectral function at high temperatures.

\begin{figure}
\includegraphics[width=8.3cm]{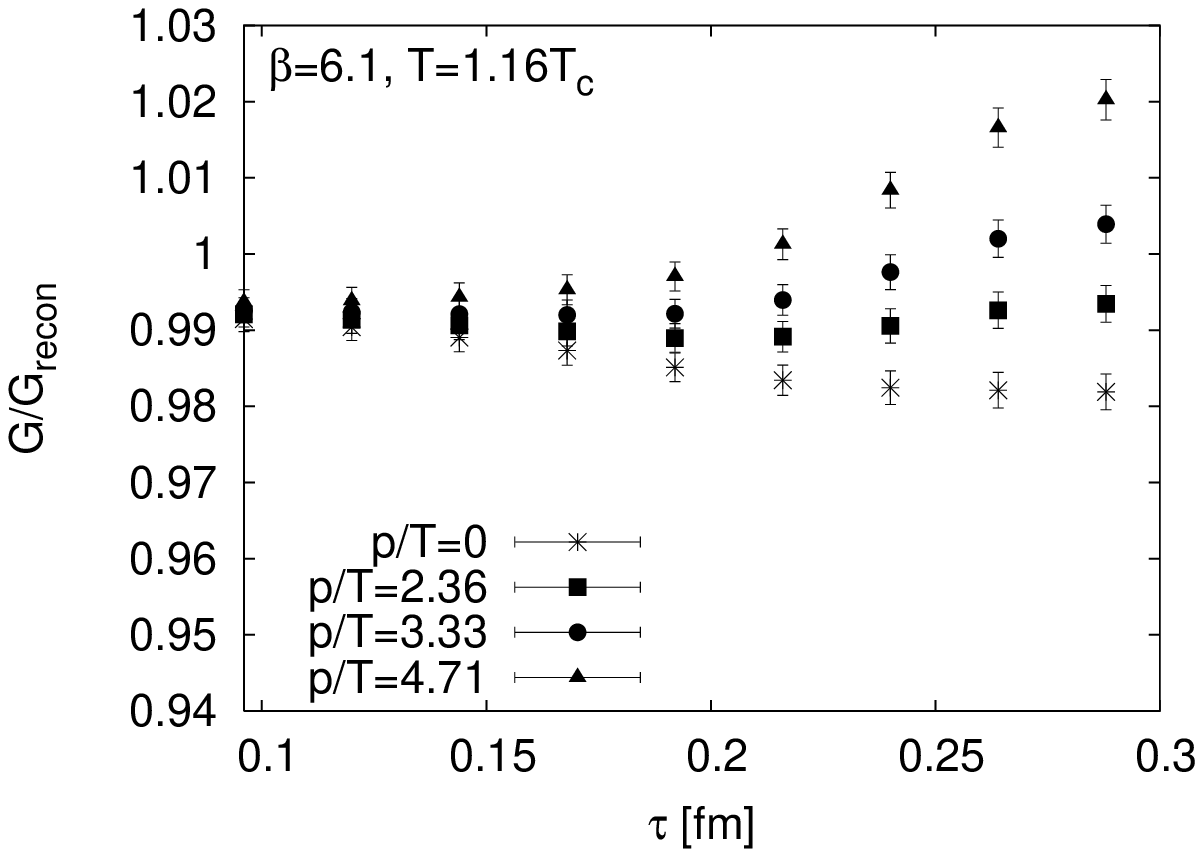}
\includegraphics[width=8.3cm]{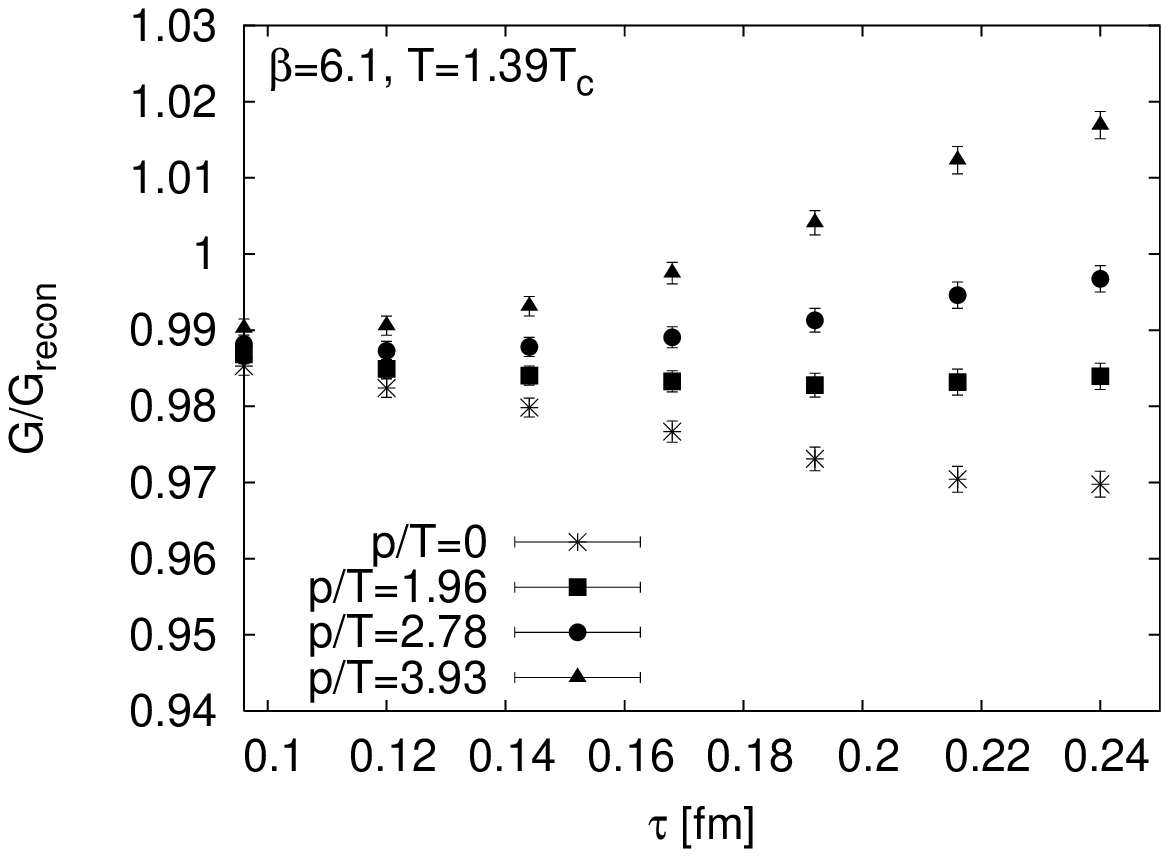}
\includegraphics[width=8.3cm]{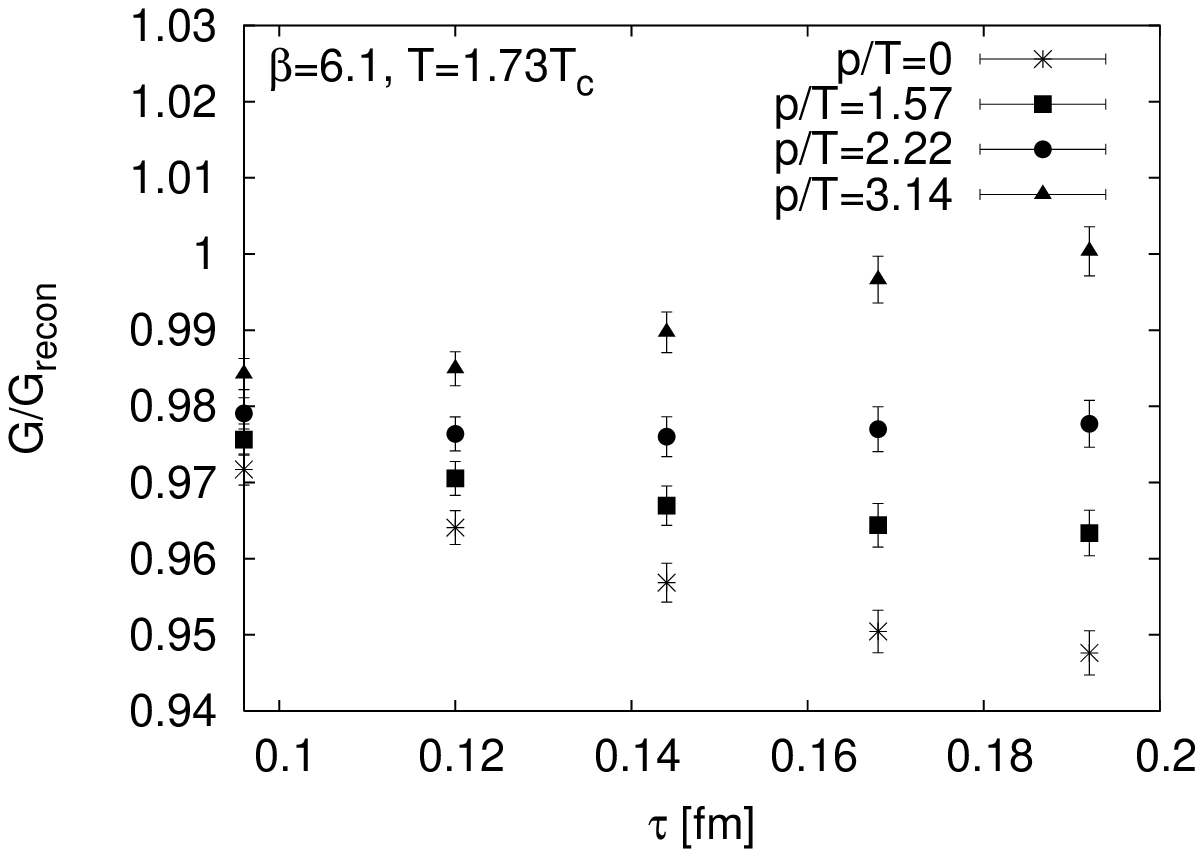}
\caption{Pseudo-scalar correlators at finite spatial momenta
calculated at three different temperatures at $\beta=6.1$ and $\xi=4$.}
\label{fig:fin_mom}
\end{figure}

\section{Charmonium correlators and spectral functions at finite momenta}
So far we considered charmonia at zero spatial momentum, i.e. charmonia at rest in the
heatbath's rest frame. It is certainly of interest to study the temperature dependence of
correlators and spectral functions at non-zero spatial momentum. Such a study has been
done using isotropic lattices with lattice spacing $a^{-1}=4.86$GeV and $9.72$ GeV \cite{datta04,datta_sewm04}.
It has been found that the pseudo-scalar correlators are enhanced compared to the zero temperature
correlators for non-vanishing spatial momenta. We have calculated the finite momentum pseudo-scalar correlators at
$\beta=6.1$, $\xi=4$ for different temperatures. The results are shown in Fig. \ref{fig:fin_mom}. We see that the
temperature dependence of $G/G_{recon}$ at zero and finite momentum are different. At zero momentum this ratio
decreases monotonically with increasing $\tau$ and increasing temperature. At finite momentum this is not the case, we see
that the decrease in   $G/G_{recon}$ is at least weaker, and at sufficiently large momenta we see even that this ratio increases
with $\tau$. The differences in $G/G_{recon}$ calculated in this work and in Refs. \cite{datta04,datta_sewm04} are
present already at zero momentum and are presumably due to finite lattice spacing errors. Apart from this 
the momentum dependence of the pseudo-scalar correlators is similar to the findings of Refs.  \cite{datta04,datta_sewm04}.
We also calculated the spectral function at finite temperature and found that the temperature dependence of the
pseudo-scalar spectral functions for non-zero momenta is significant. This again is in qualitative agreement with
the results based on isotropic lattice calculations. It would be interesting to see if the difference in the temperature
dependence of the correlators at zero and finite spatial momenta is due to a contribution to the spectral functions
below the light cone at finite temperature \cite{karsch03,aarts05}.

\section{Bottomonium spectral functions at zero temperature}

The use of Fermilab formulation described in the previous sections allows us to study also bottomonium for the same
range  of lattice spacings. Usually bottomonium is studied using lattice NRQCD (see e.g Ref. \cite{davies94}).
So far the only study of bottomonium within the relativistic framework
was presented in Ref. \cite{liao02}. The simulation parameters 
for bottomonium are summarized  in Table \ref{tab:botT=0par}. 
As before the lattice spacing is fixed by the Sommer scale $r_0$, assuming
$r_0=0.5$fm.  The values of $r_0$ for $\beta=5.9$ and $6.1$ are taken from Ref. \cite{klassen_unpub},
while for $\beta=6.3$ we use the extrapolation described in section \ref{sec:charm_lat}.
The bare quark mass, the bare velocity of light $\nu_t$ as well as the clover coefficients were taken from
Ref. \cite{liao02}.
Note that the lattice spacings determined from $r_0$ are about $20\%$ smaller than in Ref. \cite{liao02}
where it was determined from bottomonium $^1P_1-\overline{1S}$ mass splitting. As the result our estimates of the
$\Upsilon$ mass are smaller than those in Ref. \cite{liao02}. However, if we use the values of
the lattice spacing quoted in Ref. \cite{liao02} to calculate the physical masses we find good  agreement.

In our study we will stick to the value of the lattice spacing determined
from $r_0$ as it agrees reasonably well with the lattice spacings extracted from the charmonium $^1P_1-\overline{1S}$ splitting.  
This splitting is
apparently less sensitive to the effect of quenching. The bottomonium $^1P_1-\overline{1S}$ splitting, on the other hand,
is more sensitive to quenching errors as well as more difficult to estimate precisely on the lattice.
In Table \ref{tab:botT=0par} we also show the $\Upsilon$ masses.  
\begin{table}
\begin{tabular}{cccc}
\hline
\hline
$\beta$              &        5.9         &            6.1          &       6.3              \\[3mm]
\hline
$N_s \ti N_t$      &  $8^3 \ti 96$  &    $16^3 \ti 96$    &  $16^3 \ti 128$   \\[1mm]
$\xi$                  &   3.1390         &        3.2108         &     3.2686          \\[1mm]
$r_0/a$              &   3.179(9)      &        5.189(21)     &     6.91(2)          \\[1mm]
$a_t^{-1}$ [GeV] &  5.882(25)     &        8.181(32)     &    10.89(3)         \\[3mm]
\hline
$C_{sw}^s$        &  2.0598          &        1.9463        &     1.7680           \\[1mm]
$C_{sw}^t$         & 1.1277           &        1.0984        &     1.0505           \\[3mm]
\hline
$m_0$                &  1.120            &          0.670       &       0.494          \\[1mm]
$\nu_t$               &  1.50              &         1.573        &       1.50            \\[3mm]
\hline
$c(0)$                 &  1.026(32)      &         0.979(7)    &       1.004(19)    \\[1mm]
$M_{\Upsilon(1S)}$ [GeV] &  8.317(1)        &         8.084(2)    &       9.097(1)     \\[3mm]
\hline
$L_s$  [fm]          &  1.07              &         1.54          &       1.16            \\[1mm]
configs                &  700              &          650          &        650            \\
\hline
\hline
 \end{tabular}
\caption{Simulation parameters for bottomonium at zero temperature}
\label{tab:botT=0par}
\end{table}

\begin{figure}
\includegraphics[width=8.3cm]{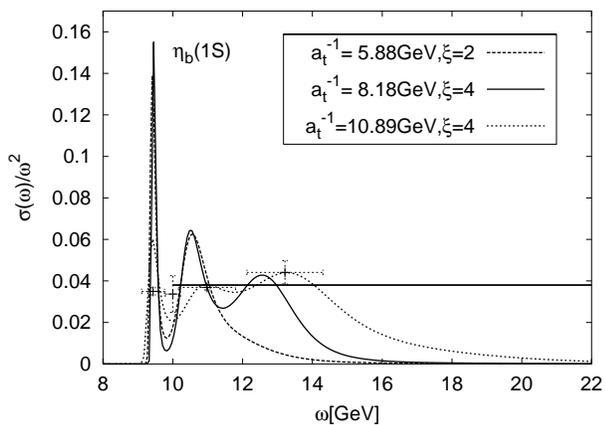}
\caption{The pseudo-scalar bottomonium spectral function at zero temperature for
different lattice spacings.}
\label{spf_ps_bot}
\end{figure}

Using MEM we analyzed the spectral functions in different channels for all three lattice spacings.
In Fig. \ref{spf_ps_bot} we show the spectral functions in the 
pseudo-scalar channel. Since the physical quark mass is different at different lattice spacings
we shifted the horizontal scale by the difference of the calculated $\Upsilon$-mass and the 
corresponding experimental value. We can see that the first peak in the spectral function corresponds to
the $\eta_b(1S)$ state and its position is independent of the lattice spacing. The remaining details
of the spectral functions are cut-off dependent and we cannot distinguish the excited states
from the continuum. 
The position and the amplitude of the first peak in the spectral functions is in good agreement with
the results of simple exponential fit. As in the charmonium case the maximal energy $\omega_{max}$ for which the spectral function is 
non-zero scales approximately as $a_s^{-1}$.  Similar results have been obtained in the vector channel.

\begin{figure}
\includegraphics[width=8.3cm]{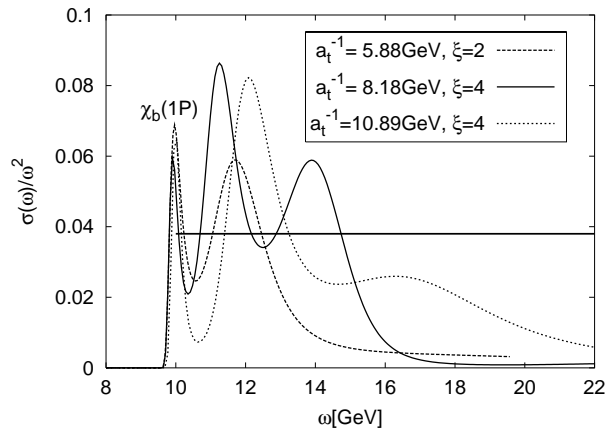}
\caption{The scalar bottomonium spectral function at zero temperature for
different lattice spacings.}
\label{spf_sc_bot}
\end{figure}

The spectral function in the scalar channel is shown in Fig.  \ref{spf_sc_bot}.
As it was the case for charmonium the correlators in this channel are more noisy
than in the pseudo-scalar channel and as a result it is more difficult to reconstruct
the spectral function. Nevertheless we are able to reconstruct the $\chi_{b0}$ state
which is the first peak in the spectral function. The peak position and the amplitude
are in reasonable agreement with the result of simple exponential fit.

\section{Bottomonia at finite temperature}
Having calculated the bottomonium spectral function at 
zero temperature we can study the temperature dependence of bottomonium
correlators to see medium modification of bottomonia properties.
The parameters of the finite temperature simulations are summarized in
Table \ref{tab:bot_ftpar}.
\begin{table*}
\begin{tabular}{|c|c|c|c|c|c|c|c|c|}
\hline
\multicolumn{3}{|c}{$\beta=5.9$ } & \multicolumn{3}{|c}{$\beta=6.1$}& 
\multicolumn{3}{|c|}{$\beta=6.3$}\\
\hline
lattice         &  $T/T_c$    & \#config &   lattice           &  $T/T_c$    &  \#config   & lattice           &  $T/T_c$      & \#config \\
$8^3 \times 16$ &    1.25     &  700     &   $16^3 \times 24$  &   1.16      &      500    & $16^3 \times 32$  &    1.15       &  440 \\
$8^3 \times 12$ &    1.66     &  700     &   $16^3 \times 20$  &   1.39      &      500    & $16^3 \times 24$  &    1.54       &  100 \\
$8^3 \times 8$  &    1.66     &  1000    &   $16^3 \times 16$  &   1.73      &      500    & $16^3 \times 20$  &    1.85       &  100 \\
$8^3 \times 6$  &    3.32     &  1000    &                     &             &             & $16^3 \times 16$  &    2.31       &  100  \\
\hline
\end{tabular}
\caption{Simulation parameters for bottomonium at finite temperature. We assumed $r_0 T_c=0.7498(50)$ corresponding to $T_c=295(2)$ MeV. }
\label{tab:bot_ftpar}
\end{table*}

\begin{figure}
\includegraphics[width=8.3cm]{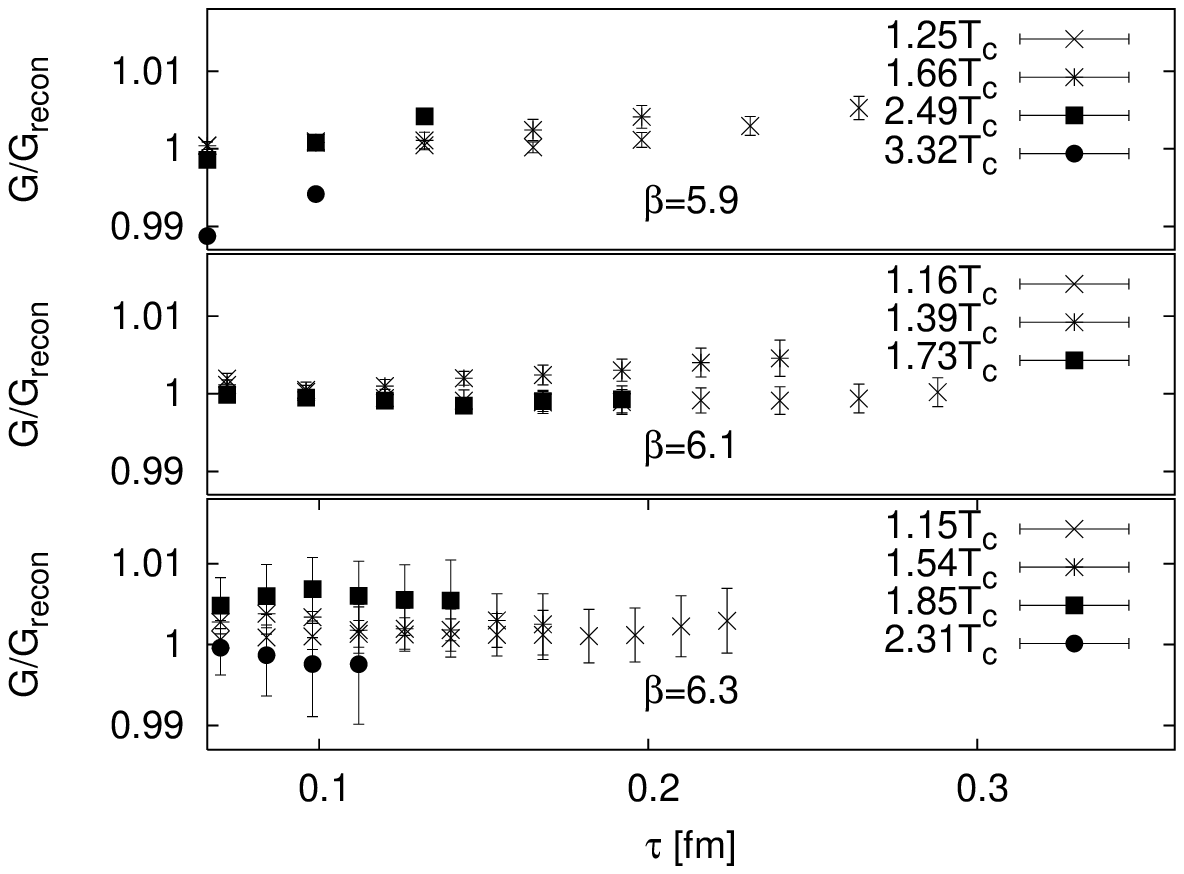}
\includegraphics[width=8.3cm]{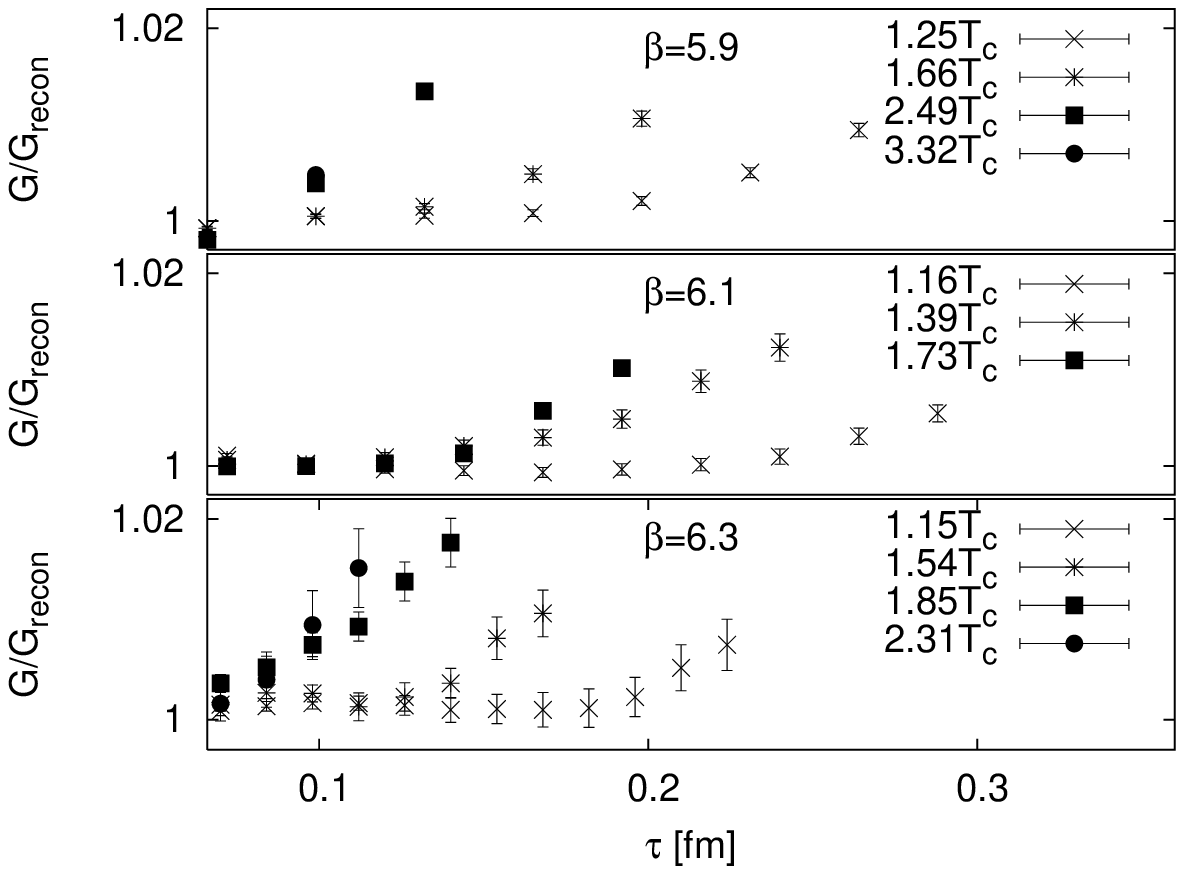}
\caption{The ratio  $G/G_{rec}$ in the pseudo-scalar (top) and vector channels at different lattice
spacings.}
\label{bot_1s}
\end{figure}

In Fig. \ref{bot_1s} we show $G/G_{recon }$ for vector and pseudo-scalar channel at different lattice spacings. 
This ratio appears to be temperature independent and very close to unity up to quite high temperatures.
This is consistent with the expectation that 1S bottomonia are smaller than 1S charmonia and thus less
effected by the medium. They could survive till significantly higher temperatures.
Compared to charmonium case the  difference between the pseudo-scalar and vector channels
is smaller. This
is also expected as the transport contribution which is responsible for this difference is proportional to $\chi_s(T)\sim \exp(-m_{c,b}/T)$,
and thus is much smaller for bottom quarks.

\begin{figure}
\includegraphics[width=8cm]{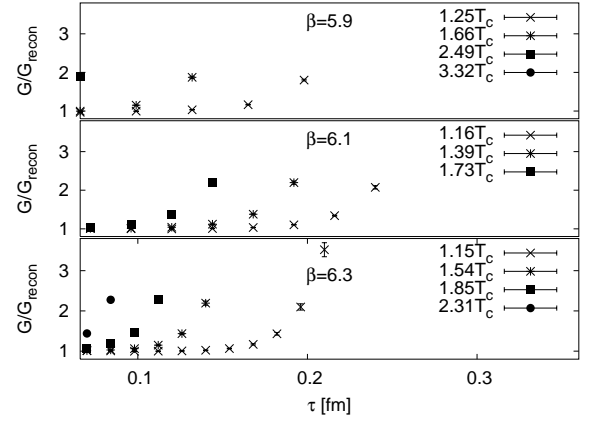}
\caption{The ratio  $G/G_{rec}$ in the scalar channel for different lattice spacings.}
\label{bot_1p}
\end{figure}

The temperature dependence of the scalar correlator is shown in Fig. \ref{bot_1p}. Contrary to the pseudo-scalar and vector 
correlators it shows strong temperature dependence and $G/G_{recon}$ is significantly larger than unity already at $1.1T_c$.  
Since the size and
the binding energy of 1P bottomonium states are comparable to that of the 1S charmonium states one would expect a much smaller
temperature dependence. The data on the other hand shows temperature dependence which is even larger than for
1P charmonium states. This may suggest that the 1P bottomonium states are strongly modified or maybe even
dissolved at $T \simeq 1.1-1.2T_c$.
Further investigation are needed to clarify what happens to 1P bottomonium states above the deconfinement
transition. Model studies of the P-wave bottomonium correlators suggest that its large change may
be due to the modification of the continuum part of the spectral functions and not to the dissolution or
strong modification of the ground state \cite{mocsy06}.  

\begin{figure}
\includegraphics[width=8cm]{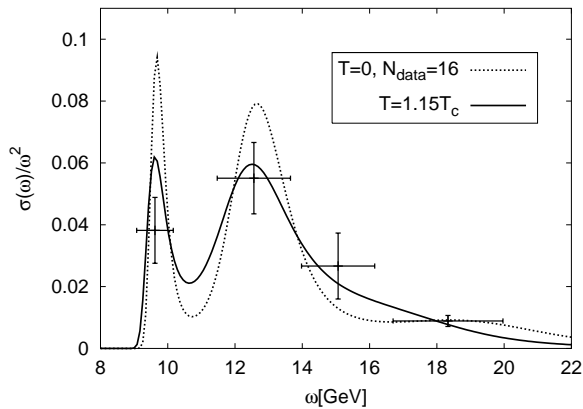}
\caption{The pseudo-scalar bottomonium spectral function at finite temperature.}
\label{spf_bot_ft}
\end{figure}

We have tried to reconstruct the spectral functions at finite temperature for $\beta=6.3$. For the pseudo-scalar channel
the results are shown in Fig. \ref{spf_bot_ft}. As in the charmonium case we compare the finite temperature spectral function
with the zero temperature spectral function obtained with the same number of data points and time interval.  As expected
the spectral function shows no temperature dependence within errors.  On the other hand we could not reliably reconstruct the
scalar spectral function at finite temperature, the probability $P[\alpha|Dm]$ showed no maximum as the function of $\alpha$.
Presumably much more statistics is needed to get some information about the scalar spectral function.

\section{Conclusions}

In this paper we investigated quarkonium correlators at zero and finite temperature in quenched QCD
using anisotropic lattices and the Fermilab formalism for the heavy quarks. 
In our investigations we used a variety of different lattice spacings 
to control cut-off effects in the quarkonium correlators and spectral functions.
Using the Maximum Entropy Method
we have calculated the spectral functions both at zero and finite temperatures. 
This was done using a newly developed algorithm for the MEM analysis which does not rely on singular value 
decomposition. It appeared to be more stable than the Bryan algorithm used so far in the
analysis of the meson spectral functions. Therefore we could for the first time reliably calculate
the charmonium spectral functions at zero temperature identifying the main structures : ground state,
excited states and the continuum. Using the zero temperature results as the base line we investigated
the temperature dependence of charmonium correlators at finite temperature at different
lattice spacings and showed that it is qualitatively the same for all lattice spacings. 
We studied the
charmonium spectral functions at finite temperature paying attention to systematic effects arising
from the fact that the time interval at finite temperature as well as the number of data points
decreases with increasing temperature. We found that the spectral functions in the 
pseudo-scalar channel do not change up to temperatures $1.5T_c$ within systematic and statistical
errors of our calculations. 
We have also learned that it is difficult to make statements about the survival of
charmonium states based solely on the shape of the spectral function as it is distorted 
by the limited time extent and strongly depends on the default model. On the other hand 
the small temperature dependence of the pseudo-scalar spectral function up to $1.5T_c$ 
holds for all default models.
The correlation functions in all channels corresponding to 1P states show very different
behavior. Together with the results for the scalar spectral functions this suggests 
the melting of the 1P charmonium states at temperature $T=1.1-1.2T_c$. 
In general, most of our findings in charmonium sector is in reasonable agreement with  results
known  from studies on fine isotropic lattices \cite{datta04}. We also identify the transport contribution in the 
vector correlators for the first time.

Our investigations in the bottomonium sector are less detailed. We were able to reconstruct the
bottomonium spectral functions at zero temperature and show that the first peak is not affected
by lattice artifacts.
Nonetheless
the spectral functions are less accurate than for charmonium.
We were not able to reconstruct reliably the bottomonium
spectral functions at finite temperature. The reason is that because of the large quark mass the
correlators decay fast and the signal is poor.   The temperature dependence of the correlators 
suggests that the 1S bottomonium state can survive till much higher temperature than the charmonium states.
On the other hand we see, unexpectedly,  drastic changes in the 1P bottomonium correlators right after 
deconfinement transition. To clarify the fate of the 1P bottomonium states further 
studies are clearly needed. They presumably should rely on NRQCD where the large bottom quark mass
is integrated out.

The use of anisotropic lattices
and Fermilab approach reduces discretization effects and allows to study the problem of quarkonium
dissolution on coarser lattices. In fact the analysis of the temperature dependence of the correlation functions
shows that they can be reliably studied already for lattice spacings $a_s^{-1}=2.0$GeV. 
Although the temperature dependence of the correlators turn out to be the same for all lattice spacings
studied in this paper, we clearly see quantitative differences. This is important when we want to extend the 
studies of quarkonium spectral functions to full QCD, where the lattice spacing is main limiting factor.
Preliminary results from calculations of charmonium spectral functions in full QCD have been reported in Ref. \cite{swan}.
These results are in qualitative agreement with findings in quenched QCD. However, no quantitative analysis
has been performed  so far to estimate the effect of  see quarks. To get sufficiently close to the continuum
limit in the analysis of the correlation function the spatial lattice spacing should presumably be smaller
than $(4 {\rm GeV})^{-1}$, i.e. it should be in the range used in  studies with isotropic lattices \cite{datta04}.

\section*{Acknowledgments} 
This work has been supported 
by U.S. Department of Energy under Contract No. DE-AC02-98CH10886 and by SciDAC project. A.V. 
was partially supported by NSF-PHY-0309362. 
K.P. is supported by Marie Curie Excellence Grant under contract MEXT-CT-2004-01
3510.
A.J. is supported by Hungarian Science Fund OTKA (F043465).
Support through LDRD funds at Brookhaven National Laboratory
is also greatly appreciated. Most of numerical calculations have been done on QCDOC supercomputer at RIKEN-BNL
Research Center. Authors used Columbia Physics System (CPS) with high-performance clover inverter by P.~Boyle
and other parts by RBC collaboration. Special thanks to C.~Jung for his generous help with CPS.
We would like to thank F. Karsch and \'A M\'ocsy for careful reading of the manuscript and valuable comments.
The final analysis of the lattice data has been done during the stay of one of the authors (P.P.) at the Institute
for Nuclear Theory (INT) at Washington University. We are grateful to INT for their support.

\section*{Appendix}

In this appendix we present some technical details on the calculations of the quarkonium spectral functions.
To study the reliability of MEM at zero temperature compare the masses and amplitudes
of different states obtained from MEM and simple 2-exponential fit. This comparison is shown in Table  
\ref{tab:compfit}. 
We also tried to estimate the systematic errors in the fit by allowing higher excited states
(3 exponential fit). We show this systematic uncertainty in the square brackets. 
One can see that we have a good agreement for the ground state in pseudo-scalar channel ($\eta_c$),
the small disagreement in the values of the meson masses (typically appearing in the 4th digit) is the consequence of the
discretization in $\omega$. We have checked that for sufficiently fine binning of the spectral function this remaining error
could be removed. 
For the scalar channel we find agreement between the results of the fit and MEM within estimated errors, with 
the only exception being $\beta=6.1$ where we find small deviation for the amplitudes which are outside the errors.
In the pseudo-scalar channel we find also an agreement for the masses and amplitudes of the excited states within
the estimated errors.   
\begin{table*}[ht]
\begin{tabular}{|c|c|c|c|c|c|c|c|c|c|c|}
\hline
&
\multicolumn{2}{|c}{$\beta=5.7$, $\xi=2$} & \multicolumn{2}{|c}{$\beta=5.9$, $\xi=2$}&
\multicolumn{2}{|c}{$\beta=6.1$, $\xi=2$ } & \multicolumn{2}{|c}{$\beta=6.1$, $\xi=4$}& 
\multicolumn{2}{|c|}{$\beta=6.5$, $\xi=4$}\\
&MEM&2-exp &MEM&2-exp &MEM&2-exp &MEM&2-exp &MEM&2-exp \\
\hline
$A_{sc}(n=1)$&0.41(2)&0.42(3)&0.211(19)&0.215(8)[9]&0.136(10)&0.111(7)[4]&0.037(6)&0.056(6)[1]&0.028(4)&0.021(1)[8]\\
$m_{sc}(n=1)$&1.832(8)&1.85(1)&1.224(45)&1.218(8)[2]&0.842(6)&0.840(4)[1]&0.410(3)&0.420(3)[4]&0.255(5)&0.251(1)[5]\\
\hline
$A_{ps}(n=1)$&1.83(2)&1.81(1)[2]&0.711(10)&0.714(2)[6]&0.32(11)&0.327(5)[2]&0.195(6)&0.1760(1)[102]&0.042(2)&0.0423(5)\\
$m_{ps}(n=1)$&1.5859(0)&1.5870(4)[3]&1.0472(0)&1.0482(1)[2]&0.727(7)&0.7285(4)[2]&0.363(5)&0.3630(3)[11]&0.2147(0)&0.2154(2)\\
\hline
$A_{ps}(n=2)$&5.22(7)&4.07(7)[88]&1.33(13)&1.8(1)[5]&0.72(31)&0.93(5)[14]&0.61(5)&0.69(36)[21]&0.107(13)&0.117(7)\\
$m_{ps}(n=2)$&2.15(4)&2.018(5)[44]&1.30(2)&1.34(1)[3]&1.01(8)&0.967(9)[14]&0.50(2)&0.481(15)[6]&0.281(8)&0.285(2)\\
\hline
\end{tabular}
\caption{The zero temperature charmonium masses and amplitudes obtained from MEM
and compared to 2-exponential fit
with jackknife errors. We show the systematic errors of the fit procedure in the square brackets.}
\label{tab:compfit}
\end{table*}

To compare the spectral functions at different lattice spacings one needs the renormalization
constants of different quark bilinears. For the lattice fermions used in the present paper these have not
been calculated. Therefore we introduced ad-hoc renormalization constants which are shown in Table
\ref{tab:z}.
\begin{table}
\begin{tabular}{|c|c|c|c|}
\hline
$\beta, ~\xi$       &   6.1,~2   &   6.1,~4    & 6.5,~4 \\
\hline
$Z_{PS}$          &  0.374    &  0.447      & 0.447  \\
$Z_{VC}$         &  0.343    &  0.447      & 0.426  \\
$Z_{SC}$          & 0.534    &   0.791      &  0.620 \\
$Z_{AX}$         &  0.592    &  0.949      &  0.707 \\
\hline
\end{tabular}
\caption{The ad-hoc renormalization constants for charmonium used in Figs. 2, 3, 5-7, 15-19
in different channels.}
\label{tab:z}
\end{table}

\begin{figure}
\includegraphics[width=7.5cm]{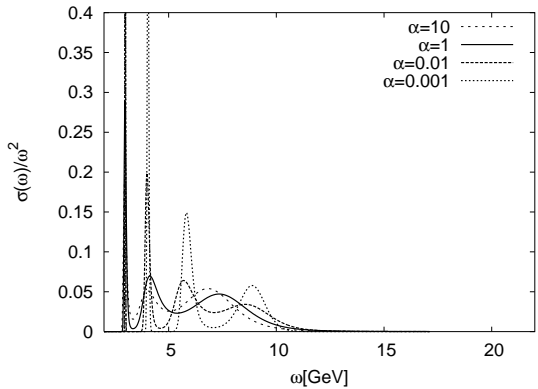}
\caption{The pseudo-scalar spectral function for different values of $\alpha$. 
The spectral functions were calculated on $16^3 \times 96$ lattice at $\beta=6.1$, $\xi=4$.}
\label{fig:alpha_dep}
\end{figure}

\begin{figure}
\includegraphics[width=7.5cm]{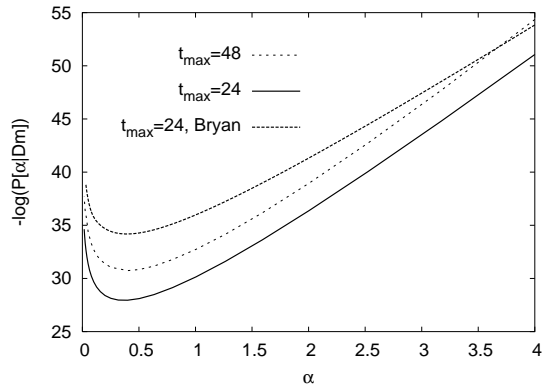}
\caption{The conditional probability
$P[\alpha|Dm]$ for different algorithms and different $t_{max}$ corresponding 
to the pseudo-scalar spectral function calculated on $16^3 \times 96$ lattice at $\beta=6.1$, $\xi=4$.}
\label{fig:Palpha}
\end{figure}

\begin{figure}
\includegraphics[width=7.5cm]{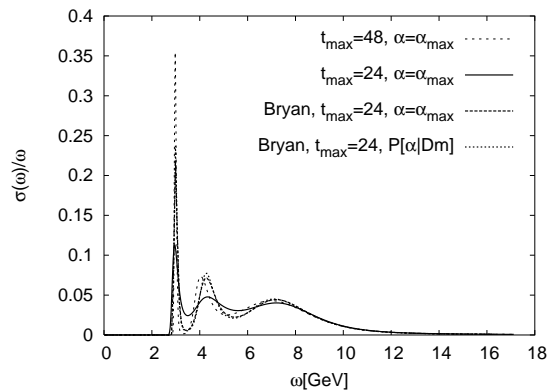}
\caption{The pseudo-scalar spectral function calculated on $16^3 \times 96$ lattice at $\beta=6.1$,
$\xi=4$ using the algorithm described in this paper and the Bryan algorithm for $\alpha=\alpha_{max}$. Also shown
in the figure is the spectral function obtained by integrating over $\alpha$ with the weight
$P[\alpha|Dm]$. See the main text for further details.}     
\label{fig:compbryan}
\end{figure}

The real parameter $\alpha$ enters into MEM through Eq. (\ref{eq:PDH}). 
Therefore we have studied the dependence of the spectral functions on $\alpha$.
For the pseudo-scalar channel at zero temperature this is shown in Fig. \ref{fig:alpha_dep}.
To get rid of the alpha dependence of the spectral function one calculates the conditional
probability $P[\alpha|DH]$. For sufficiently good quality Monte-Carlo data this function has
a unique maximum at some alpha.  This is shown in Fig. \ref{fig:Palpha} for the new 
algorithm used in this paper as well as for the Bryan algorithm.
The spectral functions calculated with the new algorithm and the Bryan algorithm are
shown in Fig. \ref{fig:compbryan}.

\begin{figure}
\includegraphics[width=7.5cm]{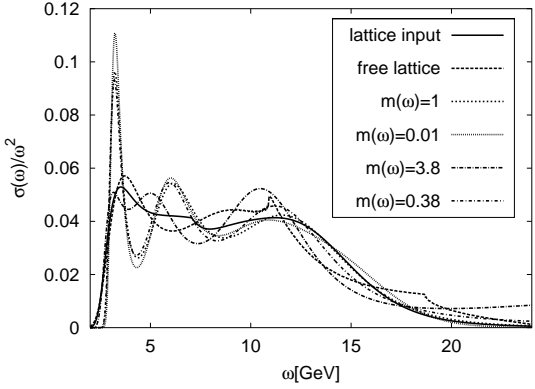}
\caption{
The default model dependence of the pseudo-scalar spectral function at $1.2T_c$ calculated on
$24^3 \times 40$ lattice with $\beta=6.5$.
}    
\label{fig:defmdep}
\end{figure}

\begin{figure}
\includegraphics[width=7cm]{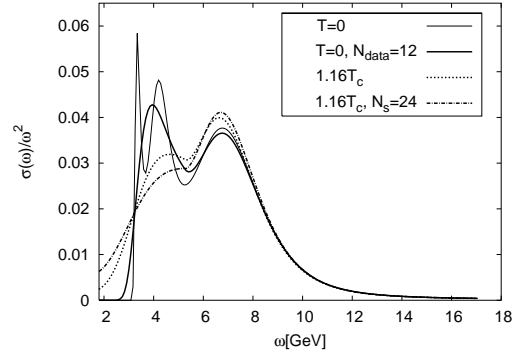}
\caption{
The scalar spectral function at $1.16T_c$ and at zero temperature reconstructed using $\tau_{max}=12$ calculated with the
default model coming from the high energy part of the zero temperature spectral function (see text). Also shown the
zero temperature spectral function calculated using all data points and $m(\omega)=1$.
}    
\label{fig:scdefamax} 
\end{figure}

\begin{figure*}
\includegraphics[width=7.7cm]{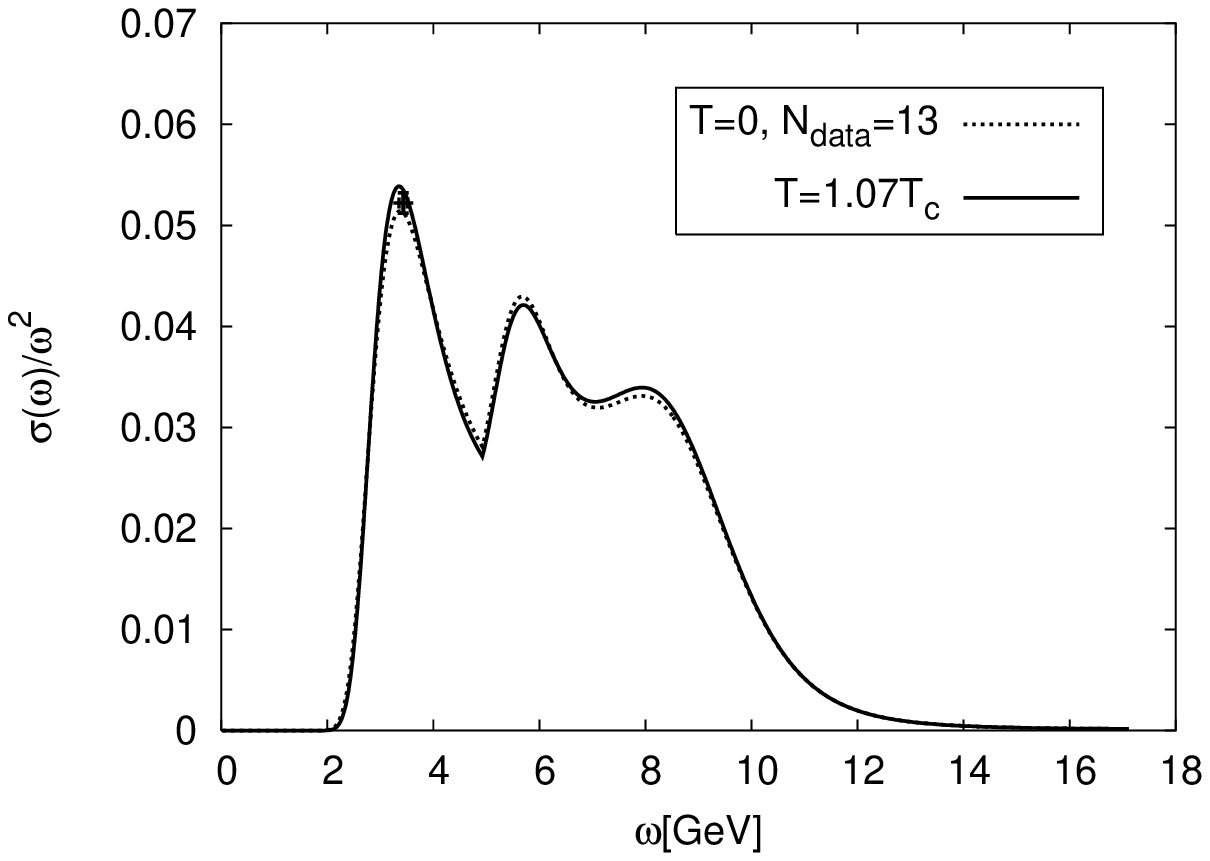}  \includegraphics[width=7.7cm]{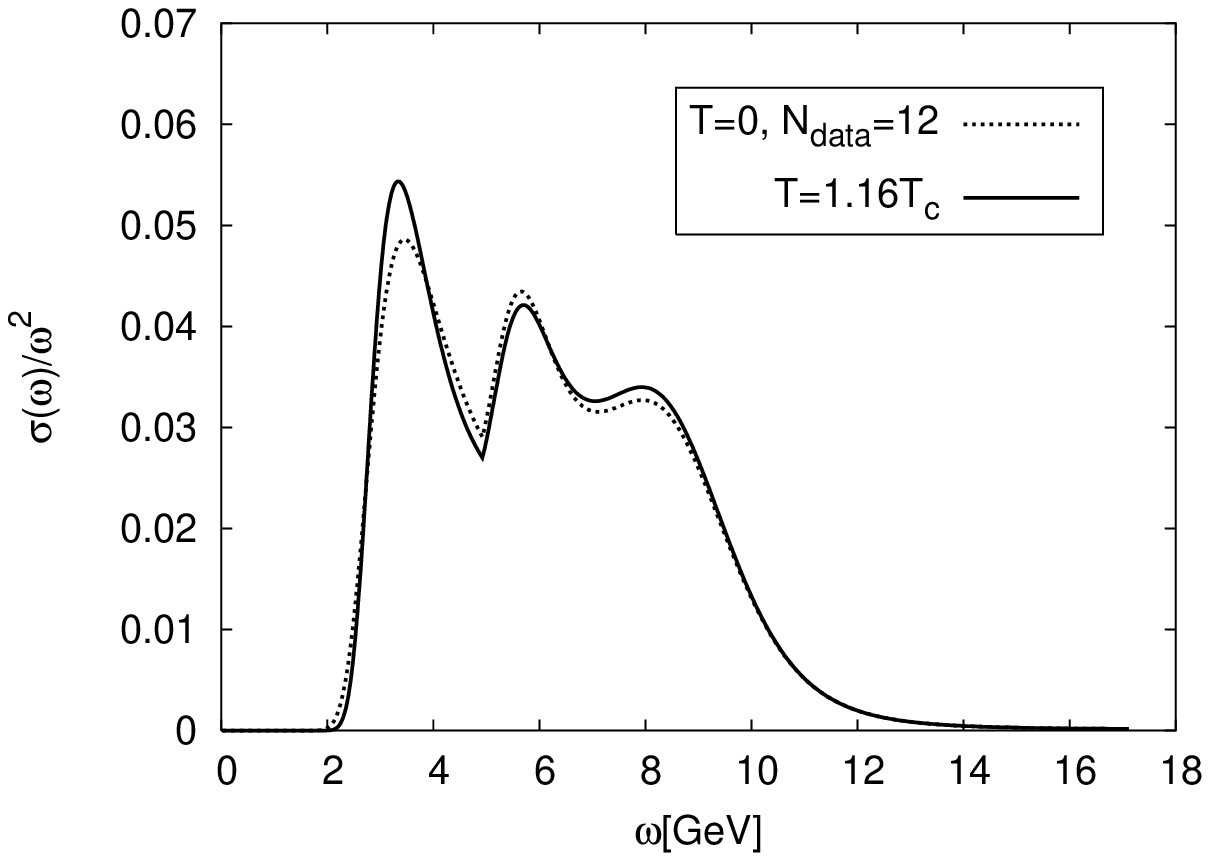}
\includegraphics[width=7.7cm]{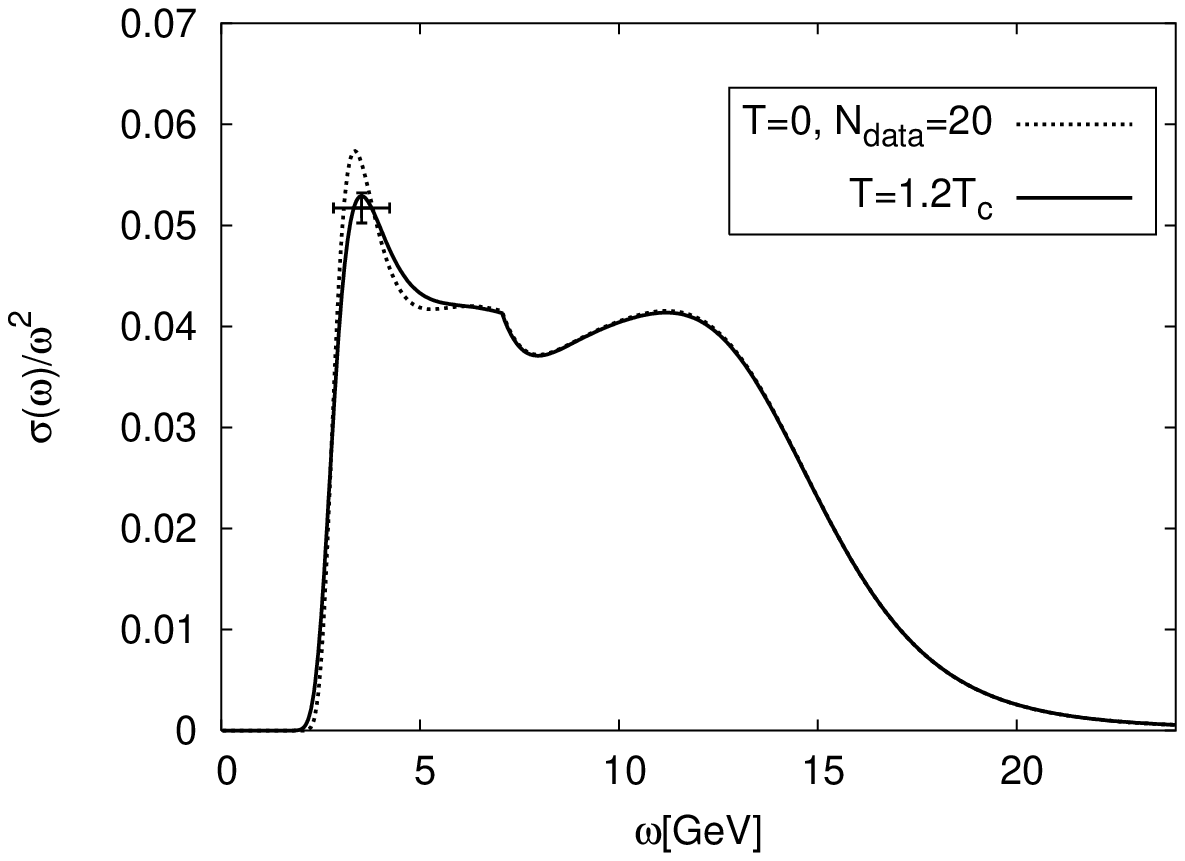}  \includegraphics[width=7.7cm]{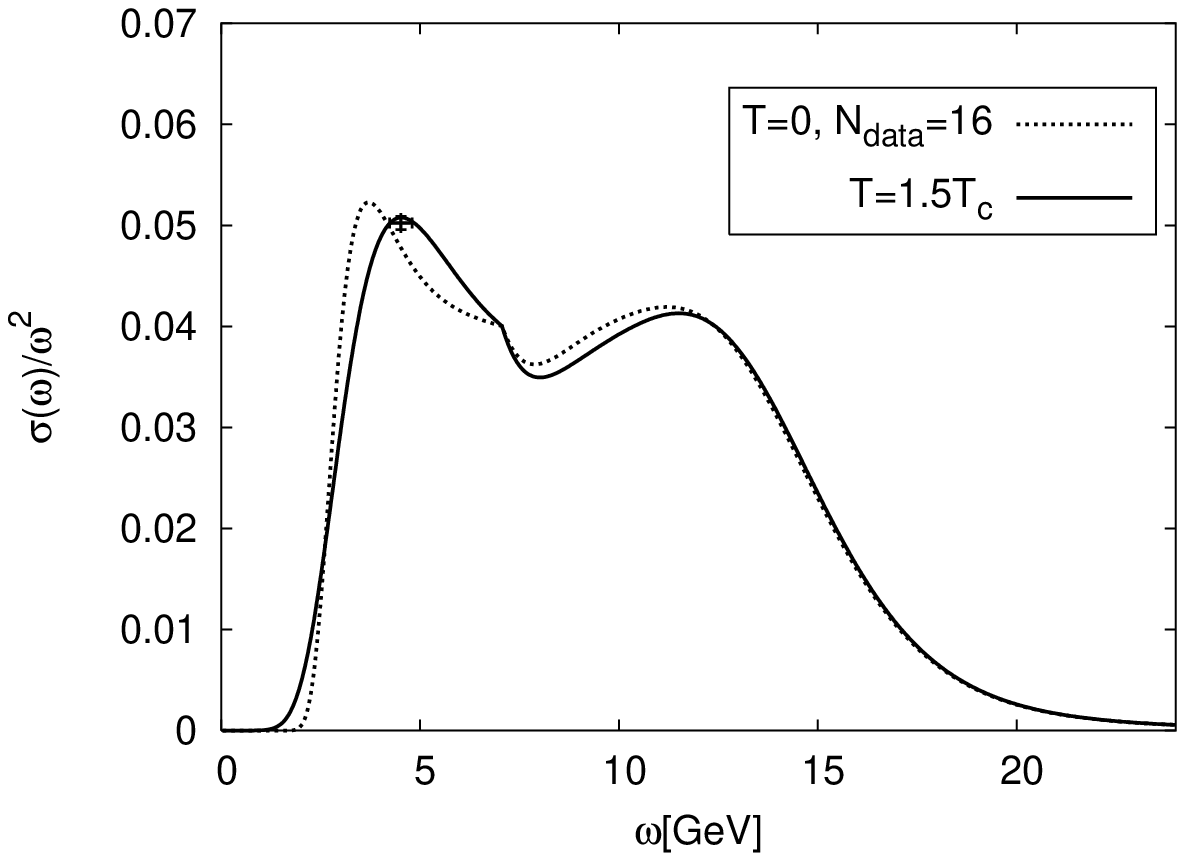}
\includegraphics[width=7.7cm]{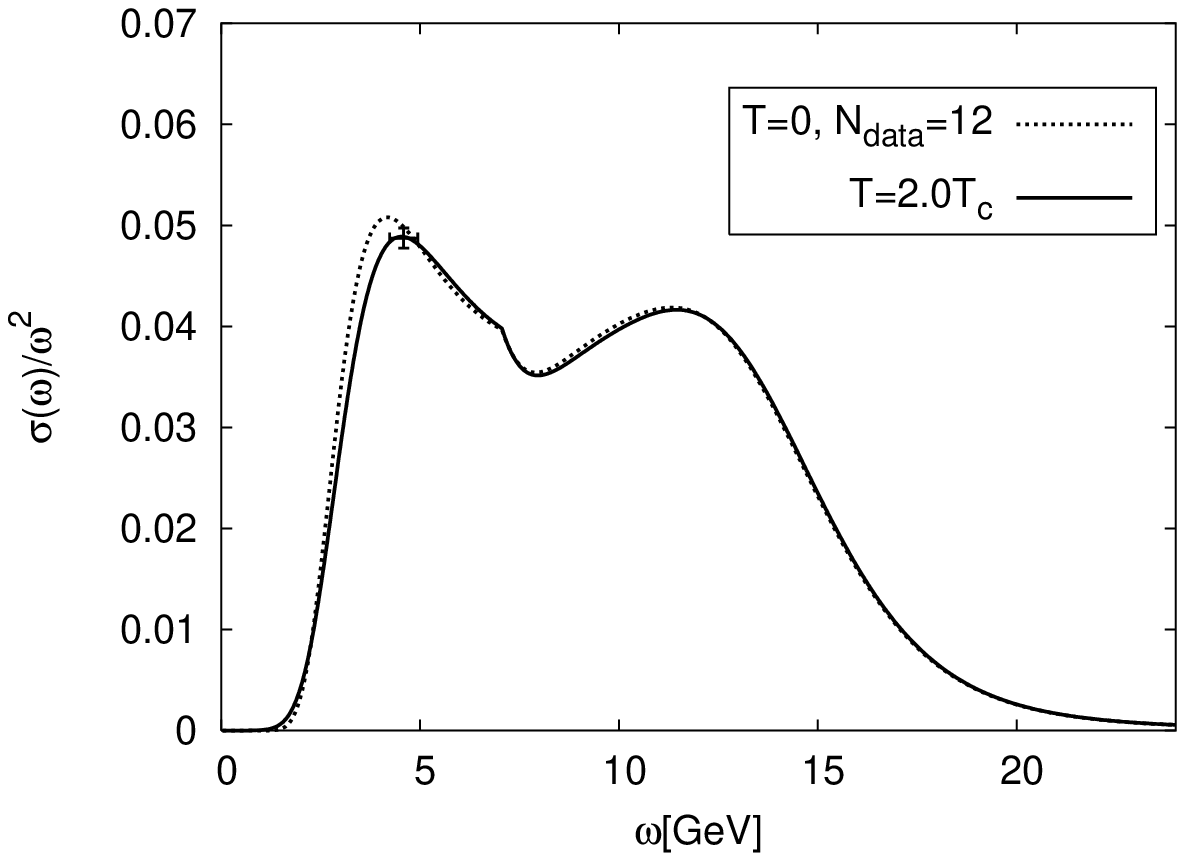} \includegraphics[width=7.7cm]{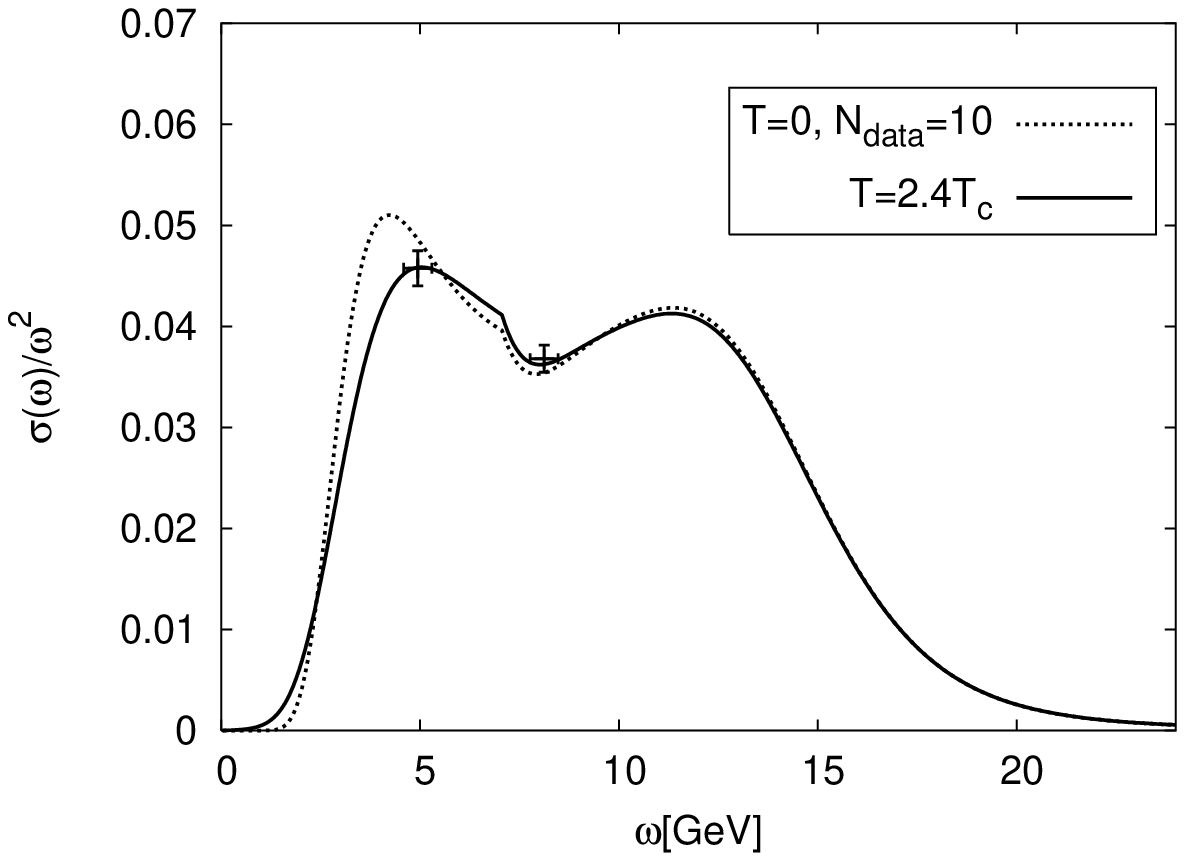}
\caption{
The pseudo-scalar spectral function at different temperatures together with the zero temperature spectral functions 
reconstructed using 
default model coming from the high energy part of the zero temperature spectral function (see text). 
}    
\label{fig:psdefamax} 
\end{figure*}

At finite temperature the dependence of the spectral functions on the default model 
becomes important. Therefore we investigated the default model dependence of the spectral function.
First we used different functional forms for the default model, $m(\omega) \sim \omega^2$ motivated by the
free spectral functions of the continuum QCD and $m(\omega)=const$. We also considered the free lattice
spectral function calculated in Ref. \cite{karsch03} as a default model. In Fig. \ref{fig:defmdep} we show
the pseudo-scalar spectral function at $1.2T_c$ calculated with different default models. The figures shows
that the default model dependence of the spectral function is significant, in particular the first peak is
not present for all default models.
Following Ref. \cite{datta04} we have used the high energy part of the zero temperature lattice spectral function
and matched it to the $\omega^2$ behavior at some energy $\omega_{th}$ above 
the energy region where no individual resonances can be seen.
In the pseudo-scalar channel at $\beta=6.5$ we have chosen $\omega_{th} \simeq 7$GeV, for  $\beta=6.1$ we 
have chosen $\omega_{th} \simeq 5$GeV in the pseudo-scalar channel and $\omega_{th} \simeq 5.5$ GeV in the scalar channel.
In Fig. \ref{fig:scdefamax} we show the scalar spectral function at zero temperature with $N_{data}=12$ and at $T=1.16T_c$
for this type of the default model.  The scalar spectral function shows significant change with the temperature in this case too.
We analyzed the pseudo-scalar spectral function with this type of the default model and the result of this analysis is shown
in Fig. \ref{fig:psdefamax}.  For this choice of the default model significant temperature dependence can be seen only for
$T=2.4T_c$. In the figure we also show the errors where appropriate. We find that the statistical errors are much smaller
for this default model.


\newpage

\begin{thebibliography}{99}

\bibitem{MS86}
T.~Matsui and H.~Satz,
Phys.\ Lett.\ B {\bf 178}, 416 (1986).

\bibitem{karsch88}
F.~Karsch, M.~T.~Mehr and H.~Satz,
Z.\ Phys.\ C {\bf 37}, 617 (1988).

\bibitem{ropke88}
G. R{\"o}pke, D. Blaschke, H. Schulz, Phys.\ Rev.\ D {\bf 38}, 3589 (1988)

\bibitem{hashimoto88}
T.  Hashimoto et al., Z.\ Phys.\ C {\bf 38}, 251 (1988)

\bibitem{digal01a}
S.~Digal, P.~Petreczky and H.~Satz,
Phys.\ Lett.\ B {\bf 514}, 57 (2001)

\bibitem{digal01b}
S.~Digal, P.~Petreczky and H.~Satz,
Phys.\ Rev.\ D {\bf 64}, 094015 (2001)


\bibitem{shuryak04}
E.~V.~Shuryak and I.~Zahed,
Phys.\ Rev.\ D {\bf 70}, 054507 (2004)


\bibitem{wong04}
  C.~Y.~Wong,
  Phys.\ Rev.\ C {\bf 72}, 034906 (2005)

\bibitem{wong06}
  C.~Y.~Wong and H.~W.~Crater,
  arXiv:hep-ph/0610440.

\bibitem{mocsy05} 
  A.~Mocsy and P.~Petreczky,
  Eur.\ Phys.\ J.\ C {\bf 43}, 77 (2005)


\bibitem{mocsy06}
  A.~Mocsy and P.~Petreczky,
  Phys.\ Rev.\ D {\bf 73}, 074007 (2006)

\bibitem{mocsy06proc}
 A.~Mocsy, P.~Petreczky and J.~Casalderrey-Solana,
 arXiv:hep-ph/0609205;
A.~Mocsy and P.~Petreczky,
arXiv:hep-ph/0606053.



\bibitem{alberico}
  W.~M.~Alberico, A.~Beraudo, A.~De Pace and A.~Molinari,
  Phys.\ Rev.\ D {\bf 72}, 114011 (2005)

\bibitem{rapp}
  D.~Cabrera and R.~Rapp,
  arXiv:hep-ph/0610254.


\bibitem{umeda02}
T.~Umeda, K.~Nomura and H.~Matsufuru,
arXiv:hep-lat/0211003.

\bibitem{asakawa04}
M.~Asakawa and T.~Hatsuda,
Phys.\ Rev.\ Lett.\  {\bf 92}, 012001 (2004)

\bibitem{datta04}
S.~Datta, F.~Karsch, P.~Petreczky and I.~Wetzorke,
Phys.\ Rev.\ D {\bf 69}, 094507 (2004) 


\bibitem{kostya_lat05}
  K.~Petrov, A.~Jakov\'ac, P.~Petreczky and A.~Velytsky,
  PoS {\bf LAT2005}, 153 (2005)
  [arXiv:hep-lat/0509138].

\bibitem{peter_lat05}
  P.~Petreczky, K.~Petrov, D.~Teaney and A.~Velytsky,
  PoS {\bf LAT2005}, 185 (2005)
  [arXiv:hep-lat/0510021].

\bibitem{jhw05}
  A.~Jakov\'ac, P.~Petreczky, K.~Petrov and A.~Velytsky,
  arXiv:hep-lat/0603005.

\bibitem{vel_hard06}
  A.~Velytsky,
  arXiv:hep-lat/0609013.

\bibitem{lebellac}
M. Le Bellac, {\em Thermal Field Theory }, Cambridge University Press ,1996

\bibitem{braaten90}
  E.~Braaten, R.~D.~Pisarski and T.~C.~Yuan,
  Phys.\ Rev.\ Lett.\  {\bf 64}, 2242 (1990).

\bibitem{karsch03}
  F.~Karsch, E.~Laermann, P.~Petreczky and S.~Stickan,
  Phys.\ Rev.\ D {\bf 68}, 014504 (2003)


\bibitem{aarts05}
  G.~Aarts and J.~M.~Martinez Resco,
  Nucl.\ Phys.\ B {\bf 726}, 93 (2005)


\bibitem{asakawa01}
  M.~Asakawa, T.~Hatsuda and Y.~Nakahara,
  Prog.\ Part.\ Nucl.\ Phys.\  {\bf 46}, 459 (2001)
  [arXiv:hep-lat/0011040].

\bibitem{lepagelat01}
  G.~P.~Lepage, B.~Clark, C.~T.~H.~Davies, K.~Hornbostel, P.~B.~Mackenzie, C.~Morningstar and H.~Trottier,
  Nucl.\ Phys.\ Proc.\ Suppl.\  {\bf 106}, 12 (2002)
  [arXiv:hep-lat/0110175].

\bibitem{jarrell96}
M. Jarrell and J.M. Gubernatis, Phys. Rep. {\bf 269} (1996) 133

\bibitem{bryan}
R. K. Bryan, Eur. Biophys J. {\bf 18}, 165 (1990)

\bibitem{nakahara99}
  Y.~Nakahara, M.~Asakawa and T.~Hatsuda,
  Phys.\ Rev.\ D {\bf 60}, 091503 (1999)
  [arXiv:hep-lat/9905034].


\bibitem{yamazaki02}
  T.~Yamazaki {\it et al.}  [CP-PACS Collaboration],
  Phys.\ Rev.\ D {\bf 65}, 014501 (2002)

\bibitem{karsch02}
  F.~Karsch, E.~Laermann, P.~Petreczky, S.~Stickan and I.~Wetzorke,
  Phys.\ Lett.\ B {\bf 530}, 147 (2002)

\bibitem{karschqm02}
  F.~Karsch, S.~Datta, E.~Laermann, P.~Petreczky, S.~Stickan and I.~Wetzorke,
  Nucl.\ Phys.\ A {\bf 715}, 701 (2003)
  [arXiv:hep-ph/0209028].

\bibitem{dattalat02}
  S.~Datta, F.~Karsch, P.~Petreczky and I.~Wetzorke,
  Nucl.\ Phys.\ Proc.\ Suppl.\  {\bf 119}, 487 (2003)
  [arXiv:hep-lat/0208012].


\bibitem{asakawalat02}
  M.~Asakawa, T.~Hatsuda and Y.~Nakahara,
  Nucl.\ Phys.\ A {\bf 715}, 863 (2003)
  [Nucl.\ Phys.\ Proc.\ Suppl.\  {\bf 119}, 481 (2003)]
  [arXiv:hep-lat/0208059].


\bibitem{blumlat04}
  T.~Blum and P.~Petreczky,
  arXiv:hep-lat/0408045.

\bibitem{mysqm03}
  P.~Petreczky,
  J.\ Phys.\ G {\bf 30}, S431 (2004)

\bibitem{mylat01}
  P.~Petreczky, F.~Karsch, E.~Laermann, S.~Stickan and I.~Wetzorke,
  Nucl.\ Phys.\ Proc.\ Suppl.\  {\bf 106}, 513 (2002)

\bibitem{hands}
  C.~R.~Allton, J.~E.~Clowser, S.~J.~Hands, J.~B.~Kogut and C.~G.~Strouthos,
  Phys.\ Rev.\ D {\bf 66}, 094511 (2002)


\bibitem{tueb}
  K.~Langfeld, H.~Reinhardt and J.~Gattnar,
  Nucl.\ Phys.\ B {\bf 621}, 131 (2002)

\bibitem{fiebig}
  H.~R.~Fiebig,
  Phys.\ Rev.\ D {\bf 65}, 094512 (2002)

\bibitem{sasaki}
  K.~Sasaki, S.~Sasaki and T.~Hatsuda,
  Phys.\ Lett.\ B {\bf 623}, 208 (2005)



\bibitem{chen01}
P. Chen, Phys.  Rev. D {\bf 64}, 034509 (2001)

\bibitem{klassen}
T. R.  Klassen, Nucl. Phys. B {\bf 533}, 557 (1998) 

\bibitem{sommer}
R. Sommer, Nucl. Phys. B {\bf 411}, 839 (1994)

\bibitem{klassen_unpub}
R. G. Edwards,  U. M.  Heller, T. R. Klassen, unpublished


\bibitem{alton94}
  C.~R.~Allton,
  Nucl.\ Phys.\ Proc.\ Suppl.\  {\bf 53}, 867 (1997)
  [arXiv:hep-lat/9610014]; hep-lat/9610016

\bibitem{okamoto02}
  M.~Okamoto {\it et al.}  [CP-PACS Collaboration],
  Phys.\ Rev.\ D {\bf 65}, 094508 (2002)
  [arXiv:hep-lat/0112020].

\bibitem{gray}
A. Gray et al, Phys. Rev. D {\bf 72}, 094507 (2005)

\bibitem{necco}
  S.~Necco,
  Nucl.\ Phys.\ B {\bf 683}, 137 (2004)
  [arXiv:hep-lat/0309017].

\bibitem{liao02}
  X.~Liao and T.~Manke,
  Phys.\ Rev.\ D {\bf 65}, 074508 (2002)
  [arXiv:hep-lat/0111049].

\bibitem{derek}
  P.~Petreczky and D.~Teaney,
  Phys.\ Rev.\ D {\bf 73}, 014508 (2006)
  [arXiv:hep-ph/0507318].

\bibitem{doi}
  H.~Iida, T.~Doi, N.~Ishii, H.~Suganuma and K.~Tsumura,
  arXiv:hep-lat/0602008.


\bibitem{datta_sewm04}
  S.~Datta, F.~Karsch, S.~Wissel, P.~Petreczky and I.~Wetzorke,
  arXiv:hep-lat/0409147.


\bibitem{davies94}
C.T. Davies, et al, Phys. Rev. D {\bf 50}, 6963 (1994)

\bibitem{swan}
  G.~Aarts, C.~R.~Allton, R.~Morrin, A.~P.~O.~Cais, M.~B.~Oktay, M.~J.~Peardon and J.~I.~Skullerud,
  arXiv:hep-lat/0610065.



\end{thebibliography}
\end{document}